\documentclass[twocolumn,trackchanges]{aastex631}

\usepackage{graphicx}	
\usepackage{rotating}
\usepackage{amsmath}	
\usepackage{amssymb}	
\usepackage{xspace}	
\usepackage{color}
\usepackage{tablefootnote}
\usepackage{fleqn,natbib}
\usepackage{threeparttable}
\usepackage{xcolor}
\usepackage{dcolumn}
\usepackage{float}
\usepackage{multirow,longtable,supertabular}
\usepackage{hyperref}

\newcommand{\thisgrb}{GRB~230204B\xspace}

\newcommand{\fermi}{{\em Fermi}\xspace}

\newcommand{\fermiT}{{T$_{\rm 0}$}\xspace}
\newcommand{\keV}{{\rm keV}\xspace}

\newcommand{\s}{{\rm ~s}\xspace}
\newcommand{\swift}{{\em Swift}\xspace}
\newcommand{\tninty}{{$T_{\rm 90}$}\xspace}

\newcommand{\Ep}{$E_{\rm p}$\xspace}
\newcommand{\sw}[1]{\texttt{#1}}

\received{December, 2024}
\accepted{\today}

\graphicspath{{./}{figures/}}

\begin{document}

\title{Extremely luminous optical afterglow of an energetic gamma-ray burst GRB 230204B}

\correspondingauthor{Rahul Gupta}
\email{rahulbhu.c157@gmail.com, rahul.gupta@nasa.gov}

\author[0000-0003-4905-7801]{Rahul Gupta}
\affiliation{Astrophysics Science Division, NASA Goddard Space Flight Center, Mail Code 661, Greenbelt, MD 20771, USA}
\affiliation{NASA Postdoctoral Program Fellow}

\author[0000-0002-4744-9898]{Judith Racusin}
\affiliation{Astrophysics Science Division, NASA Goddard Space Flight Center, Mail Code 661, Greenbelt, MD 20771, USA}

\author[0000-0003-3668-1314]{Vladimir Lipunov}
\affiliation{Lomonosov Moscow State University, 119234, Moscow, Universitetskiy prospect, 13, 37, Russia} 

\author[0000-0002-7400-4608]{Y.-D. Hu}
{\affiliation{INAF-Osservatorio Astronomico di Brera, Via E. Bianchi 46, 23807 Merate, LC, Italy}
\affiliation{Instituto de Astrof\'isica de Andaluc\'ia (IAA-CSIC), Glorieta de la Astronom\'ia s/n, E-18008, Granada, Spain}

\author[0000-0001-8802-520X]{Ashna Gulati}
\affiliation{Sydney Institute for Astronomy, School of Physics, The University of Sydney, NSW 2006, Australia}
\affiliation{ARC Centre of Excellence for Gravitational Wave Discovery (OzGrav), Hawthorn, VIC 3122, Australia}
\affiliation{CSIRO Space and Astronomy, PO Box 76, Epping, NSW 1710, Australia}

\author[0000-0003-2999-3563]{Alberto J. Castro-Tirado}
\affiliation{Instituto de Astrof\'isica de Andaluc\'ia (IAA-CSIC), Glorieta de la Astronom\'ia s/n, E-18008, Granada, Spain}
\affiliation{Ingenier\'ia de Sistemas y Autom\'atica, Universidad de M\'alaga, Unidad Asociada al CSIC por
el IAA, Escuela de Ingenier\'ias Industriales, Arquitecto Francisco
Pe\~nalosa, 6, Campanillas, 29071 M\'alaga, Spain}

\author[0000-0002-2686-438X]{Tara Murphy}
\affiliation{Sydney Institute for Astronomy, School of Physics, The University of Sydney, NSW 2006, Australia}
\affiliation{ARC Centre of Excellence for Gravitational Wave Discovery (OzGrav), Hawthorn, VIC 3122, Australia}

\author{Motoko Serino}
\affiliation{Department of Physical Sciences, Aoyama Gakuin University, 5-10-1 Fuchinobe, Chuo-ku, Sagamihara, Kanagawa 252-5258, Japan}

\author{Kirill Zhirkov}
\affiliation{Lomonosov Moscow State University, 119234, Moscow, Universitetskiy prospect, 13, 37, Russia} 

\author[0009-0009-4622-7749]{S. Shilling}
\affiliation{Department of Physics, Lancaster University, Lancaster, LA1 4YB, UK}
\affiliation{Astrophysics Science Division, NASA Goddard Space Flight Center, Mail Code 661, Greenbelt, MD 20771, USA}
\affiliation{Center for Research and Exploration in Space Science and Technology, NASA Goddard Space Flight Center, Greenbelt, MD, 20771, USA}

\author[0000-0001-9309-7873]{Samantha R. Oates}
\affiliation{Department of Physics, Lancaster University, Lancaster, LA1 4YB, UK}

\author[0000-0002-9415-3766]{James K. Leung}
\affiliation{David A. Dunlap Department of Astronomy and Astrophysics, University of Toronto, 50 St. George Street, Toronto, ON M5S 3H4, Canada}
\affiliation{Dunlap Institute for Astronomy and Astrophysics, University of Toronto, 50 St. George Street, Toronto, ON M5S 3H4, Canada}
\affiliation{Racah Institute of Physics, The Hebrew University of Jerusalem, Jerusalem 91904, Israel}

\author[0000-0002-4299-2517]{T. Parsotan}
\affiliation{Astrophysics Science Division, NASA Goddard Space Flight Center, Mail Code 661, Greenbelt, MD 20771, USA}

\author[0000-0003-3164-8056]{Amit K. Ror}
\affiliation{Aryabhatta Research Institute of Observational Sciences (ARIES), Manora Peak, Nainital-263002, India}

\author{Shashi B. Pandey}
\affiliation{Aryabhatta Research Institute of Observational Sciences (ARIES), Manora Peak, Nainital-263002, India} 

\author[0000-0003-3220-7543]{S. Iyyani}
\affiliation{Indian Institute of Science Education and Research, Thiruvananthapuram, Kerala, India, 695551}
\affiliation{High Performace Computing Centre, IISER Thiruvananthapuram, Kerala, India, 695551}

\author[0000-0002-4394-4138]{V. Sharma}
\affiliation{Astrophysics Science Division, NASA Goddard Space Flight Center, Mail Code 661, Greenbelt, MD 20771, USA}

\author[0000-0002-9928-0369]{A. Aryan}
\affiliation{Graduate Institute of Astronomy, National Central University, 300 Jhongda Road, 32001 Jhongli, Taiwan}

\author{Jin-Ming Bai} 
\affiliation{Yunnan Observatories, Chinese Academy of Sciences, Kunming 650216, China}

\author{Pavel Balanutsa}
\affiliation{Lomonosov Moscow State University, 119234, Moscow, Universitetskiy prospect, 13, 37, Russia} 

\author[0000-0002-7004-9956]{David Buckley}
\affiliation{AAA Department of Astronomy, University of Cape Town, Private Bag X3, Rondebosch 7701, South Africa} 
\affiliation{AAB South African Astronomical Observatory, PO Box 9, 7935 Observatory, Cape Town, South Africa} 

\author[0000-0001-7920-4564]{Mar\'ia D. Caballero-Garc{\'i}a}
\affiliation{Instituto de Astrof\'isica de Andaluc\'ia (IAA-CSIC), Glorieta de la Astronom\'ia s/n, E-18008, Granada, Spain}

\author{I. M. Carrasco-Garc{\'i}a}
\affiliation{Sociedad Malague\~na de Astronom{\'i}a, Rep{\'u}blica Argentina 2, 29071 M{\'a}laga, Spain}

\author{A. Castell{\'o}n}
\affiliation{Department of Algebra, Geometry and Topology, Facultad de Ciencias, Universidad de M\'alaga, Boulevard Luis Pasteur, s\/n, E-29071, M\'alaga, Spain}

\author{Sebastián Castillo}
\affiliation{Servicios Centrales de Investigaci{\'o}n, Universidad de M{\'a}laga, Boulevard Luis Pasteur, s/n, E-29071, M{\'a}laga, Spain}

\author{Chen-Zhou Cui} 
\affiliation{Beijing National Astronomical Observatory, Beijing, China}

\author{Yu-Feng Fan} 
\affiliation{Yunnan Observatories, Chinese Academy of Sciences, Kunming 650216, China}

\author[0009-0009-4604-9639]{Emilio Fern{\'a}ndez-Garc{\'\i}a}
\affiliation{Instituto de Astrof\'isica de Andaluc\'ia (IAA-CSIC), Glorieta de la Astronom\'ia s/n, E-18008, Granada, Spain}

\author{Guillermo Garc{\'i}a-Segura}
\affiliation{Instituto de Astronom\'{\i}a, Universidad Nacional Aut\'onoma de M\'exico,  Carr. Tijuana-Ensenada km.107, 22860 Ensenada, B.C., Mexico}
\affiliation{Instituto de Astrof\'isica de Andaluc\'ia (IAA-CSIC), Glorieta de la Astronom\'ia s/n, E-18008, Granada, Spain}

\author[0000-0003-4268-6277]{Maria Gritsevich}
\affiliation{Swedish Institute of Space Physics (IRF), Bengt Hultqvists v{\"{a}}g 1, 981 92 Kiruna, Sweden}
\affiliation{University of Helsinki, Faculty of Science, Gustav Hällströmin katu 2, FI-00014, Finland}
\affiliation{Institute of Physics and Technology, Ural Federal University, Mira str. 19, 620002 Ekaterinburg}

\author[0000-0003-2628-6468]{Sergiy Guziy}
\affiliation{Instituto de Astrof\'isica de Andaluc\'ia (IAA-CSIC), Glorieta de la Astronom\'ia s/n, E-18008, Granada, Spain}
\affiliation{Petro Mohyla Black Sea National University, Mykolaiv 54000, Ukraine}\affiliation{Mykolaiv Astronomical Observatory, Mykolaiv 54030, Ukraine}

\author[0000-0002-4711-7658]{David Hiriart}
\affiliation{Instituto de Astronom\'{\i}a, Universidad Nacional Aut\'onoma de M\'exico,  Carr. Tijuana-Ensenada km.107, 22860 Ensenada, B.C., Mexico}

\author[0000-0002-2467-5673]{William H. Lee}
\affiliation{Instituto de Astronom\'ia, Universidad Nacional Aut{\'o}noma de M{\'e}xico, Apdo Postal 70-264, Cd. Universitaria, 04510 M{\'e}xico DF, M{\'e}xico}

\author[0009-0007-4214-0692]{Soomin Jeong}
\affiliation{Agency for Defense Development, Daejeon 34060, Republic of Korea}

\author{Carlos Jesus P{\'e}rez del Pulgar}
\affiliation{Unidad Asociada al CSIC Departamento de Ingenier{\'i}a de Sistemas y Autom{\'a}tica, Escuela de Ingenier{\'i}as, Universidad de M{\'a}laga, C\/. Dr. Ortiz Ramos s/n, 29071 M{\'a}laga, Spain}

\author{Ignacio Olivares}
\affiliation{Instituto de Astrof\'isica de Andaluc\'ia (IAA-CSIC), Glorieta de la Astronom\'ia s/n, E-18008, Granada, Spain}

\author[0000-0001-6665-9631]{Il H. Park}
\affiliation{Department of Physics, Sungkyunkwan University, Seobu-ro 2066, Suwon, 16419 Korea}

\author[0000-0002-7273-3671]{Ignacio P{\'e}rez-Garc{\'i}a}
\affiliation{Instituto de Astrof\'isica de Andaluc\'ia (IAA-CSIC), Glorieta de la Astronom\'ia s/n, E-18008, Granada, Spain}

\author[0000-0002-0130-2460]{S.~Razzaque}
\affiliation{Centre for Astro-Particle Physics (CAPP) and Department of Physics, University of Johannesburg, PO Box 524, Auckland Park 2006, South Africa}
\affiliation{Department of Physics, The George Washington University, Washington, DC 20052, USA}
\affiliation{National Institute for Theoretical and Computational Sciences (NITheCS), Private Bag X1, Matieland, South Africa}

\author[0000-0002-7158-5099]{Rubén S{\'a}nchez-Ram{\'i}rez}
\affiliation{Instituto de Astrof\'isica de Andaluc\'ia (IAA-CSIC), Glorieta de la Astronom\'ia s/n, E-18008, Granada, Spain}

\author{Nataly Tiurina}
\affiliation{Lomonosov Moscow State University, 119234, Moscow, Universitetskiy prospect, 13, 37, Russia} 

\author{Vladislav Topolev}
\affiliation{Lomonosov Moscow State University, 119234, Moscow, Universitetskiy prospect, 13, 37, Russia} 

\author{Chuan-Jun Wang}
\affiliation{Yunnan Observatories, Chinese Academy of Sciences, Kunming 650216, China}

\author[0009-0004-7113-8258]{Si-Yu Wu}
\affiliation{Instituto de Astrof\'isica de Andaluc\'ia (IAA-CSIC), Glorieta de la Astronom\'ia s/n, E-18008, Granada, Spain}

\author{Yu-Xin Xin} 
\affiliation{Yunnan Observatories, Chinese Academy of Sciences, Kunming 650216, China}

\author[0000-0002-6809-9575]{Ding-Rong Xiong}
\affiliation{Yunnan Observatories, Chinese Academy of Sciences, Kunming 650216, China}

\author{Xiao-Hong Zhao}
\affiliation{Yunnan Observatories, Chinese Academy of Sciences, Kunming 650216, China}

\author{Jirong Mao}
\affiliation{Yunnan Observatories, Chinese Academy of Sciences, Kunming 650216, China}

\author{Bao-Li Lun}
\affiliation{Yunnan Observatories, Chinese Academy of Sciences, Kunming 650216, China}

\author{Ye Kai}
\affiliation{Yunnan Observatories, Chinese Academy of Sciences, Kunming 650216, China}

\begin{abstract}

Robotic telescope networks play an important role in capturing early and bright optical afterglows, providing critical insights into the energetics and emission mechanisms of GRBs. In this study, we analyze GRB 230204B, an exceptionally energetic and multi-pulsed long GRB, detected by the \fermi GBM and MAXI detectors, with an isotropic equivalent gamma-ray energy exceeding 10$^{54}$ erg. Time-resolved spectral analysis reveals a transition in the prompt emission from hard (sub-photospheric dominated) spectra during early pulses to softer (synchrotron radiation dominated) spectra in later pulses, indicative of a hybrid jet composition. We report the discovery and characterization of the optical afterglow using the MASTER and BOOTES robotic telescope networks, which enabled rapid follow-up observations starting at $\sim$1.3 ks post-burst. The optical luminosity at this time was exceptionally high, surpassing that of many other optically bright GRBs, such as GRB 990123, GRB 080319B, etc. This places the burst among the most luminous optical GRBs observed to date. Long-term radio observations extending to 335 days post-burst were conducted with the ATCA. Multi-wavelength modeling, incorporating data from MASTER, BOOTES, DOT, \swift/XRT, and radio observations, was conducted using an external ISM forward-shock top-hat jet model with \sw{afterglowpy}. The results reveal a narrow and highly collimated jet with a circumburst density of $n_{0} \sim$ 28.12 cm$^{-3}$, kinetic energy $E_{\rm K} \sim$ 4.18 $\times 10^{55}$ erg, and a relatively low value of $\epsilon_{B}$ = 2.14 $\times 10^{-6}$, indicating shock-compression of magnetic field in the surrounding interstellar medium. We constrained a low radiative efficiency of $\sim$ 4.3 \%. This study highlights the indispensable contribution of robotic networks to early afterglow observations and advances our understanding of GRB 230204B unique characteristics and underlying jet physics.
 
\end{abstract}

\keywords{Gamma-ray bursts}

\section{Introduction} 
\label{sec:intro}

Gamma-ray bursts (GRBs) are among the most energetic events in the Universe, emitting vast amounts of energy within a few seconds to minutes, often followed by an afterglow that can be observed across the electromagnetic spectrum \citep{1995ARA&A..33..415F, 1997ApJ...476..232M, 1998ApJ...497L..17S, 1999PhR...314..575P, 2000ApJ...543...66P, 2001ApJ...562L..55F, 2004RvMP...76.1143P, 2006RPPh...69.2259M, 2006ApJ...642..354Z, 2006ARA&A..44..507W, 2014ARA&A..52...43B, 2015PhR...561....1K, 2017ApJ...848L..12A}. The discovery of afterglows in the late 1990s marked a significant milestone in GRB research. The first detection of an X-ray afterglow by \cite{1997Natur.387..783C}, optical afterglow by \cite{1997Natur.386..686V}, and radio afterglow by \cite{1997Natur.389..261F} enabled precise localization and redshift determination. This, in turn, provided insights into the distances and energies involved in GRBs \citep{1999Natur.398..389K}. The study of GRB afterglows, particularly in the optical regime, has been pivotal in understanding the progenitor systems \citep{2001Sci...291...79M}, the environment surrounding the burst \citep{2010ApJ...720.1513K, 2012ApJ...758...27L}, and the mechanisms driving the emission \citep{2023arXiv231216265G}. The temporal evolution of these afterglows, often characterized by a power-law decay, can reveal crucial details about the jet dynamics, energy injection, and interaction with the surrounding medium \citep{1997ApJ...476..232M, 1998ApJ...497L..17S}.

In recent times, robotic telescope networks, such as Mobile Astronomical System of Telescope-Robots (MASTER, \citealt{Li2022}), Burst Observer and Optical Transient Exploring System (BOOTES, \citealt{2023NatAs...7.1136C}), Las Cumbres Observatory (LCO) Global Telescope \citep{2013PASP..125.1031B}, Rapid Eye Mount (REM, \citealt{2004SPIE.5492.1613C}), Liverpool Telescope (LT, \citealt{2016RMxAC..48...48C}), Gravitational-wave Optical Transient Observer (GOTO, \citealt{2022MNRAS.511.2405S}), Multicolor Imaging Telescopes for Survey and Monstrous Explosions (MITSuME, \citealt{2008AIPC.1000..543S}), the Robotic Optical Transient Search Experiment (ROTSE III, \citealt{2003PASP..115..132A}), Nazarbayev University Transient Telescope at Assy-Turgen Astrophysical Observatory (NUTTelA-TAO, \citealt{2022SPIE12184E..8AG}), and others, have become indispensable in the rapid follow-up of GRBs, often identifying optical afterglows within minutes to hours (especially for \fermi Gamma-ray Burst Monitor (GBM) detected GRBs with large error circles) of the initial burst \citep{2010AdAst2010E..30L, 2012ASInC...7..313C}. \fermi GBM’s localization accuracy, while adequate for high-energy studies, often leaves large localization uncertainties spanning a few to tens of degrees radius \citep{2009ApJ...702..791M, 2015ApJS..216...32C, 2020ApJ...895...40G}, making rapid and wide-field follow-up observations essential. These robotic telescope networks, equipped with wide-field cameras and fully automated systems, are specifically designed to address this challenge. Their ability to quickly respond to GRB alerts and scan large areas of the sky allows them to pinpoint transient optical emissions within the \fermi error regions. Additionally, their prompt response times enable them to observe early afterglow phases, providing data at these early times critical for informing the onset and evolution of the emission \citep{2021MNRAS.505.4086G, 2023ApJ...942...34R, 2024A&A...692A...3S}. These observations are useful for constraining GRB redshifts indirectly by identifying afterglow candidates for follow-up spectroscopic studies. Furthermore, the robotic nature of these observatories ensures continuous and autonomous temporal coverage, improving the chances of detecting GRBs in real-time and providing crucial data to guide further multi-wavelength follow-up efforts \citep{2023NatAs...7.1136C}.

GRB 230204B, detected by the \fermi GBM and MAXI, stands out as one of the most energetic GRBs observed, with an isotropic equivalent energy (E$_{\gamma, iso}$) exceeding 10$^{54}$ erg (see Section \ref{amati}). The rapid and precise localization of GRB 230204B by the MAXI detector triggered an immediate response from the MASTER and BOOTES robotic telescope networks, which provided crucial early-time optical data that significantly shaped our physical interpretation of this burst. These networks are specifically designed for the rapid follow-up of transient astronomical events, such as GRBs, and their automated observing capabilities are crucial for observing the early phases of the optical afterglow \citep{2023NatAs...7.1136C}. The MASTER network began observing GRB 230204B at $\sim$ 1.3 ks post-burst, providing the earliest optical detection of the afterglow of \thisgrb and constrained the exceptionally high optical luminosity  at the time of MASTER/BOOTES detection ($\sim$ 1.3 ks post-burst), allowing us to place GRB~230204B among the most luminous optical afterglows known. The BOOTES network complemented these observations with additional early-time data, which were essential for constraining the decay phase of the optical light curve (see Section \ref{afterglowanalyis}). In this paper, we present a detailed analysis of the optical and multi-wavelength follow-up observations (see Section \ref{sec:observations}) of GRB 230204B, including data from MASTER \citep{2023GCN.33441....1L}, BOOTES \citep{2023NatAs...7.1136C}, Devasthal Optical Telescope (DOT, \citealt{2023GCN.33284....1R, 2024BSRSL..93..683G}), \swift Ultraviolet and Optical telescope (UVOT, \citealt{2023GCN.33292....1S, 2023Univ....9..113O}), Growth India Telescope (GIT, \citealt{2023GCN.33269....1S}), Asteroid Terrestrial-impact Last Alert System (ATLAS, \citealt{2023GCN.33278....1S}), Very Large Telescope (VLT, \citealt{2023GCN.33281....1S}), Australia Telescope Compact Array (ATCA, \citealt{2023GCN.33321....1G}), and Australian Square Kilometre Array Pathfinder (ASKAP, \citealt{2020PASA...37...48M}) telescope. We explore the temporal and spectral characteristics of the prompt emission and afterglow and compare its optical luminosity with other known GRBs, the results are presented in Section \ref{sec:results}. Discussion on potential physical mechanisms driving this energetic burst and the afterglow brightness comparison are given in Section \ref{sec:discussion}. The key findings are summarized in Section \ref{sec:SUMMARY}.

\section{Observations and data analysis} 
\label{sec:observations}

This section details the observation campaign for \thisgrb, observed by multiple space-based and ground-based telescopes across a broad wavelength range. Comprehensive details of the observing campaign are provided below, and the sequence of prompt and afterglow observations for \thisgrb is shown in Figure \ref{fig:seq1}.

\begin{figure}[!ht]
\centering
\includegraphics[scale=0.36]{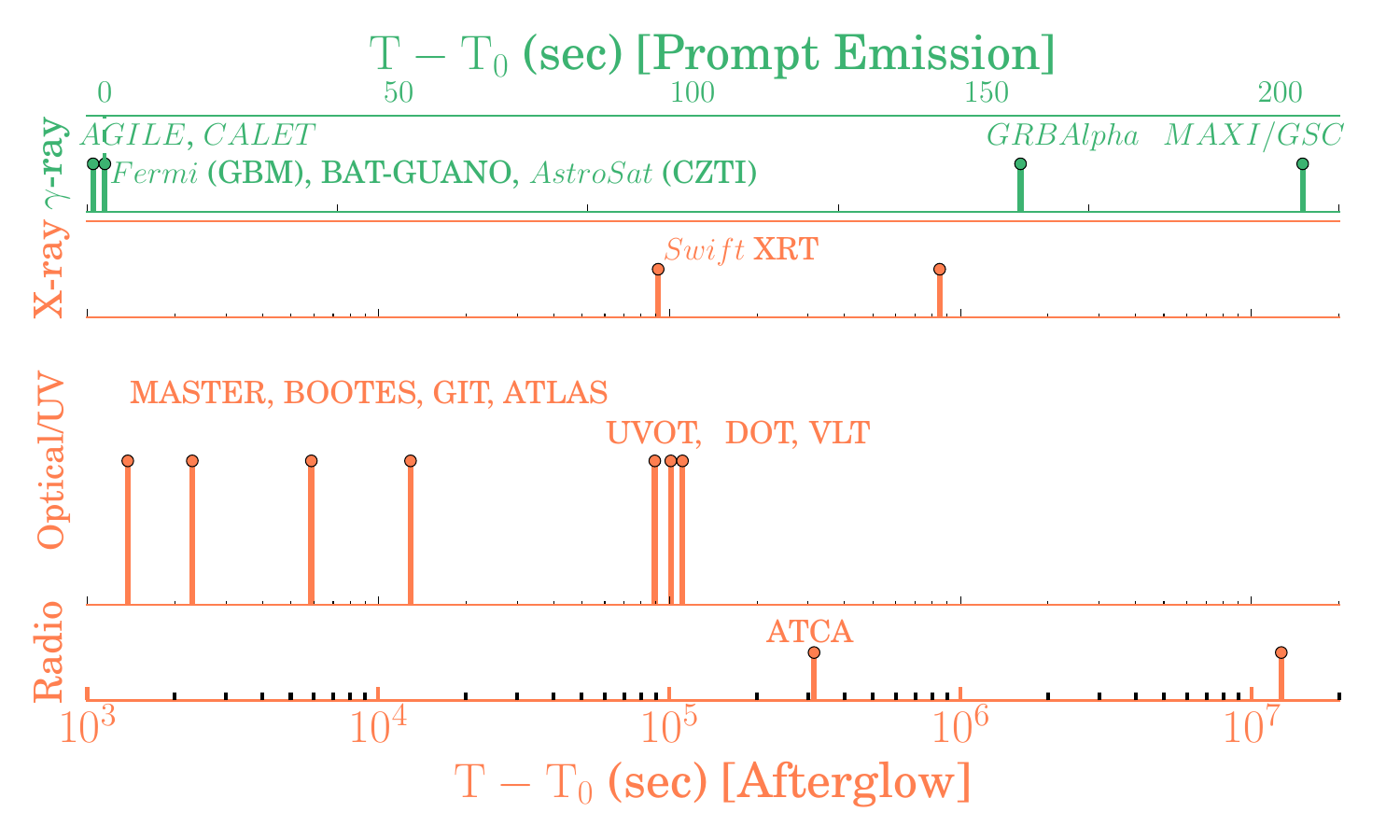} 
\caption{Sequence of prompt and afterglow observations for \thisgrb, the dashed vertical line indicating the \fermi-GBM trigger time. Vertical green and orange lines with dots correspond to the start times of observations by respective instruments.} 
\label{fig:seq1}
\end{figure}

\subsection{Prompt Emission}

GRB 230204B was first detected on 4th February 2023 by the \fermi GBM satellite at 21:44:27.20 UT (hereafter \fermiT), with \tninty duration of 216 seconds in 50-300 \keV \citep{2023GCN.33288....1P}. MAXI also detected it at 21:47:51 UT and the burst position was constrained to approximately RA = 13h 10m 19s, Dec = -21\textdegree\ 45$'$ 07$''$ (J2000) with a statistical uncertainty region of 0.1\textdegree \citep{2023GCN.33265....1S}. The burst did not trigger the Neil Gehrels \swift Observatory; henceforth \swift \citep{2004ApJ...611.1005G}. However, following the MAXI trigger notification, the Gamma-ray Urgent Archiver for Novel Opportunities (GUANO, \citealt{2020ApJ...900...35T}) pipeline requested data from the onboard buffer of the \swift Burst Alert Telescope (BAT) and initiated a search within the 200-second event mode data centered around \fermiT. The GRB was detected in this \swift/BAT GUANO data, determining its \tninty\ to be approximately 65 seconds \citep{2023GCN.33267....1K}. The burst was also detected by the Indian multiwavelength satellite \textit{AstroSat}, which measured a \tninty\ of $216_{-22.56}^{+5.13}$ seconds \citep{2023GCN.33268....1W}. The prompt emission of the burst was also seen by \textit{AGILE} duration $\sim$ 145 sec, \citealt{2023GCN.33272....1C}) and \textit{GRBAlpha} (\tninty of 207 sec, \citealt{2023GCN.33273....1D}) missions, with every mission providing a slightly different value of \tninty duration as expected due to the different sensitive and spectral coverage.

\subsection{Prompt X-ray Emission: MAXI}

MAXI (mounted on the {\it ISS}) scans a position of the sky every 92 minutes \citep{maxi}. The position of the GRB moved into the MAXI field of view at \fermiT+192.8\,sec, and the scan transit ended at \fermiT+248.8 sec. During the scan transit, the MAXI/GSC nova alert system \citep{negoro} triggered the burst at 21:47:51 UT (\fermiT+ 203.8 sec), with MAXI's detection revealing the soft nature of the burst. The detection occurred during the final stages of the GRB emission, as evidenced by its temporal proximity to the main burst and the significant variability observed in the MAXI light curve during the scan transit (see Figure \ref{fig:promptlc}). The observed temporal variability is characteristic signature of prompt emission \citep{2004IJMPA..19.2385Z, 2015PhR...561....1K, 2023A&A...671A.112C} rather than early afterglow, which typically exhibits smoother temporal evolution \citep{1998ApJ...497L..17S, 2006ApJ...642..389N, 2006ApJ...647.1213O}. These characteristics support the interpretation that the detected X-rays are part of the prompt emission of \thisgrb rather than the afterglow. Assuming the source flux was constant over the transit, the X-ray flux in the 4.0--10.0 keV band was calculated to be 628 $\pm$ 49 mCrab (1 sigma error, \citealt{2023GCN.33265....1S}).

After the detection of the late phase of prompt emission of \thisgrb by MAXI/GSC, it could not observe the position of the burst for about 6 hours due to the observation constraint of the radiation zone \citep{sugizaki} in the International Space Station (ISS) orbit. The MAXI observations resumed at $\sim$ \fermiT+ 22.5\,ks, and there were 11 consecutive scan transits to search for an X-ray afterglow of \thisgrb. The afterglow was not detected in any transit of MAXI observations. We calculated 3 sigma upper limits in the 2--20 keV band (see Table \ref{tab:MAXI} in appendix) for these transits following the method by \citet{sugita}. This non-detection by MAXI/GSC during the late phase further supports the prompt nature of the initial MAXI detection. A deeper search for the X-ray counterpart of GRB 230204B was later ($\sim$ \fermiT+ 80\,ks) carried out by \swift/XRT (see Section \ref{XRT}).

\subsection{Afterglow: Soft X-ray Emission}
\label{XRT}

The fading X-ray counterpart of GRB 230204B was detected by \swift/XRT approximately 80 ks after the burst trigger \citep{2023GCN.33285....1D}. Given the faint nature of the X-ray afterglow at such a late epoch, all observations were performed in photon counting (PC) mode to optimize sensitivity for faint sources \citep{2007A&A...469..379E, 2009MNRAS.397.1177E}. Follow-up observations were conducted during two epochs (XRT exposure times of 5.1\,ks, and 4.5\,ks, respectively), on 5 February 2023 and 14 February 2023, resulting in only two usable data points (see Table A2 of the appendix). The final observation on February 14 approached the detection limit of the instrument, emphasizing the faintness of the X-ray emission and the challenges of constraining the X-ray afterglow’s properties at these late times. 

\subsection{Afterglow: Optical Emission}

We discovered the optical afterglow of \thisgrb using two robotic telescope networks (using MASTER in the unfiltered band and using BOOTES in the clear band, respectively) independently (see Figure \ref{fig:seq2}). The optical counterpart of the burst was also identified by the 0.7m GROWTH-India Telescope (GIT) approximately 1.57 hours (later than MASTER and BOOTES) after the burst trigger \citep{2023GCN.33269....1S}. This optical afterglow was subsequently observed by several other ground-based telescopes \citep{2023GCN.33278....1S, 2023GCN.33284....1R}. Using the X-shooter spectrograph on the Very Large Telescope (VLT), \cite{2023GCN.33281....1S}  estimated a redshift of $z$ = 2.142 for the burst. The given magnitude (for the positive detection) and upper limits within 3$\sigma$ are listed in Table~\ref{tab:optical}. 

\subsubsection{MASTER robotic telescope}

The MASTER global network comprises nine identical fully robotic wide-field twin telescopes, distributed around the Earth for all-sky monitoring up to 19-20 mag. Its primary goal is to detect optical transients, including high-energy astrophysical events such as GRBs \citep{Li2010, Li2019, Li2022, Li2023}. The network operates in both open (8 square degrees FOV) and closed (4 square degrees FOV) modes, enabling rapid follow-up observations of alerts from facilities such as \textit{Fermi}, LIGO/Virgo, IceCube, ANTARES/KM3Net, \swift, MAXI, \fermi-LAT, GECAM, \textit{Einstein Probe}, and INTEGRAL \citep{Kornilov, Troja, Mundell, Gupta1}.

\begin{figure}
\centering
\includegraphics[scale=0.4]{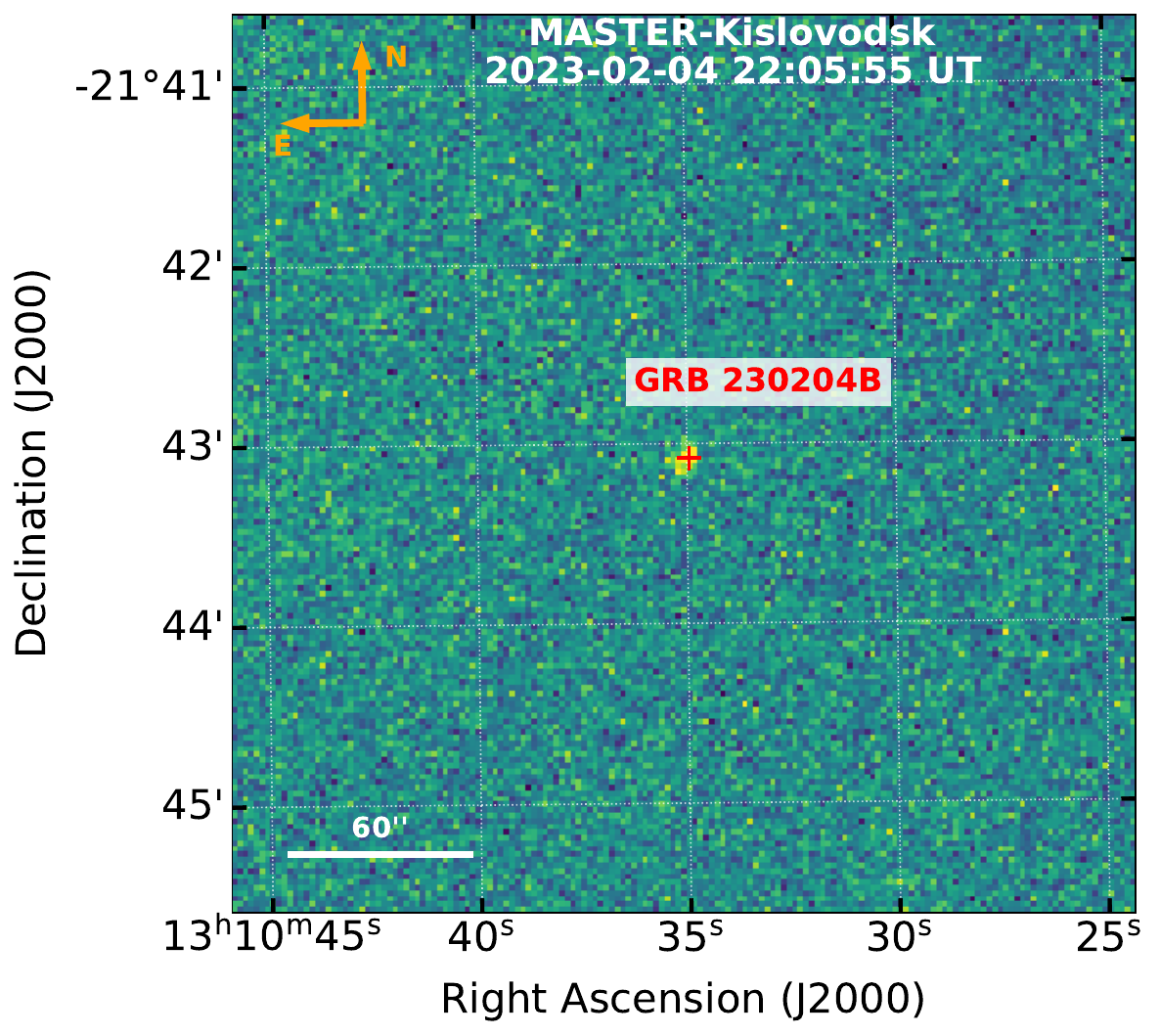}
\includegraphics[scale=0.4]{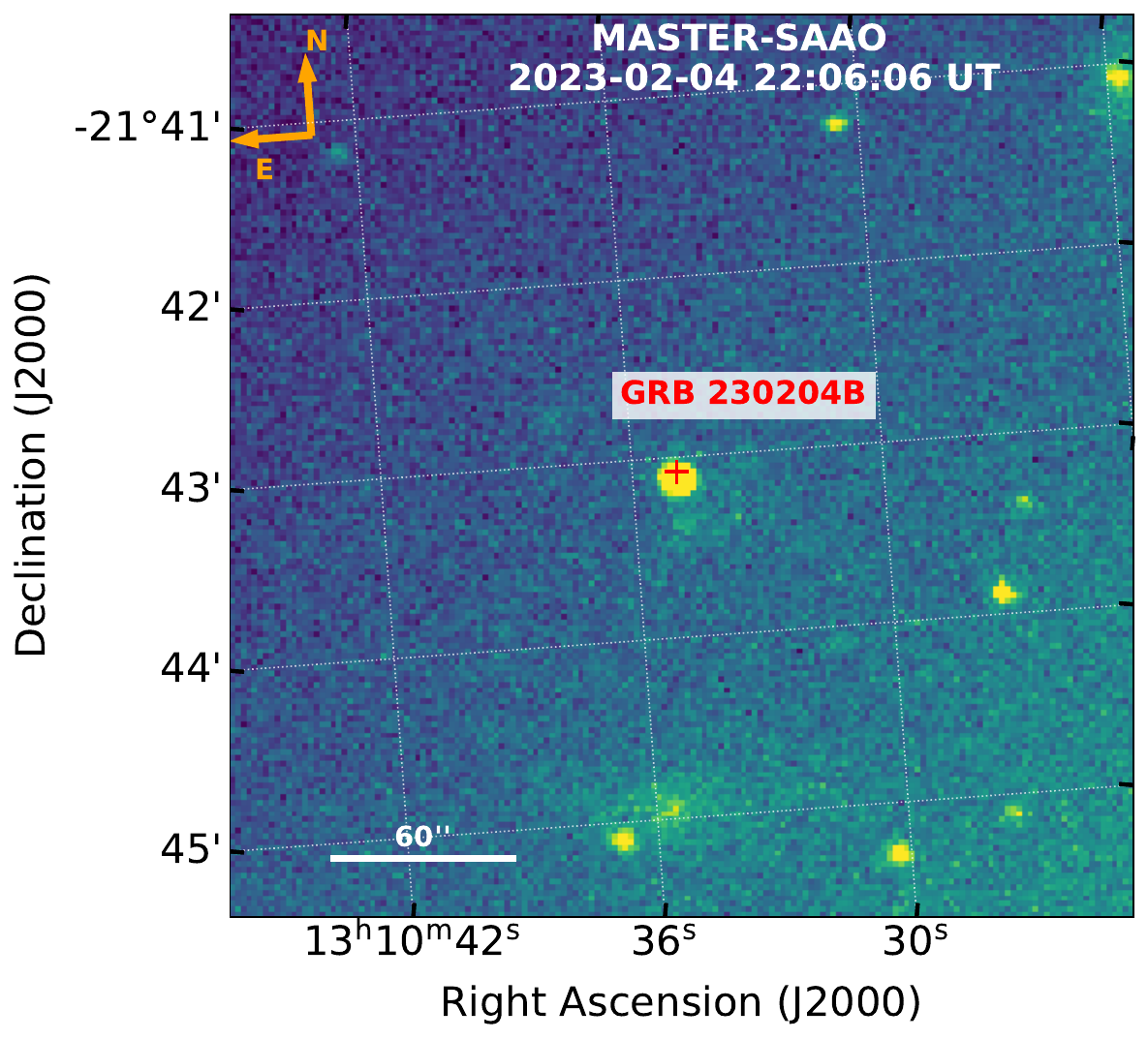}
\caption{Optical finding chart for GRB 230204B from observations taken by MASTER-Kislovodsk (Top), and MASTER-SAAO (Bottom). The afterglow position is marked with a red plus marker in each image.}
\label{fig:seq2}
\end{figure}

GRB 230204B was observed by MASTER-Kislovodsk and MASTER-SAAO (South African Astronomical Observatory). First images were received for the MAXI error box, then telescopes automatically pointed to \fermi field. The MASTER-Kislovodsk robotic telescope started observing the MAXI trigger 979708795 observation at 2023-02-04 22:05:55 UT (1064\,sec after trigger time which was 74\,sec after the notification time, see Figure \ref{fig:seq2}) at 11$^{\circ}$ altitude, in 44$^\circ$ distance to the Moon (0.3 phase, that creates significant interference for wide-field instruments observations) through clouds at horizon with $m_{lim}$=14.8 mag at first image in polarization filter. The transient was not detected in this initial image.

The MASTER-SAAO robotic telescope located in the South African Astronomical Observatory, started observing error box automatically 85 seconds after notice time (i.e. 1075\,sec after trigger time) at 2023-02-04 22:06:06 UT with optical counterpart MASTER OT J131034.94-214304.8 detection, coincident with the counterpart reported by GIT, ATLAS, VLT, DOT, and BOOTES\footnote{\url{https://gcn.gsfc.nasa.gov/other/230204B.gcn3}}. The observations began at 27$^\circ$ altitude with an exposure time of 180\,sec in the polarization filter. The OT unfiltered magnitude in the first image was 12.9 mag ($m_{lim}$=17.6 mag). We used a reference image on 2020-04-24.82 UT with unfiltered $m_{lim}$=20.6 mag.

For the photometric analysis of this object, catalog data for comparison stars were sourced from Gaia EDR3 to evaluate the variance in magnitude errors. The photometry dataset excludes frames where the comparison star data were deemed unreliable. The magnitude errors for comparison stars were computed as the deviation between the weighted average magnitude (where weights are the inverse square of the signal-to-noise ratio) and the magnitude measured on individual frames. Two approaches were employed to determine the stellar magnitude. In the first approach, a reference star from Gaia EDR3 was identified, its flux on the observational frame measured, and the magnitude calculated using Pogson's scale. This method served as a validation step to assess consistency and identify potential systematic errors. In the second approach, which was used for the final magnitude determination, a zero-point correction was applied to the reference star’s magnitude, utilizing the comparison stars as secondary references. The second method generally produced a smaller magnitude variance compared to the first, owing to the improved calibration achieved through the zero-point correction. By employing this dual-approach methodology, we ensured robust photometric accuracy while mitigating systematic errors. The final results presented in this study are based solely on the second method.
   
\subsubsection{BOOTES robotic telescope}

BOOTES followed up GRB 230204B with three of the six 60\,cm robotic telescopes, namely the ones at the BOOTES-4/MET station in Lijiang Astronomical Observatory (Yunnan, China), at BOOTES-5/JGT station in Observatorio Astron\'omico Nacional of San Pedro M\'artir (Baja California, M\'exico) and at the BOOTES-2/TELMA station in La Mayora (M\'alaga, Spain). Following the MAXI trigger, the BOOTES-4/MET telescope performed two epoch observations at 2023-02-04 22:19:12 UT and 2023-02-05 17:42:04 UT. The BOOTES-2/TELMA telescope and the BOOTES-5/JGT telescope also executed follow-up observations at 2023-02-05 04:16:55 UT and 2023-02-05 07:43:45 UT, respectively. A series of images were obtained using the clear filter with exposures of 1 sec, 5 sec, and 10\,sec at BOOTES-4/MET telescope during the first epoch, and the optical afterglow was clearly detected during the decaying phase. However, during the subsequent multi-filter monitoring (Sloan-$r$$i$$z$ and clear filters, 60\,sec exposure), the optical afterglow was not detected on the stacked images for each epoch. This non-detection is likely due to the fact that the late-time stacked images were not deep enough to detect the rapidly fading afterglow. All the images were processed with bias, dark subtraction, and flat-field correction using custom IRAF routines \citep{2021MNRAS.505.4086G}. Photometry was carried out using the standard IRAF package, and the images were calibrated with nearby comparison stars from the USNO-B1.0 (for clear filter) and SDSS (for $r$$i$$z$ filters) catalogs.

\subsubsection{DOT}

We initiated observations using the DOT \citep{2016RMxAC..48...83P, 2024BSRSL..93..683G} on 2023-02-06 at 01:49:47.939 UT, approximately 1.16 days after the trigger. Multiple frames with an exposure time of 180\,sec each were observed in the $g$, $r$, and $i$ bands, and continuous observations were conducted over four consecutive nights using the same filters. For image preprocessing, we employed the standard IRAF framework, and DAOPHOT II was used to determine magnitudes from the cleaned files. We stacked the images for each night in the same filter to increase the signal-to-noise ratio. PSF photometry was performed on the final stacked images \citep{2022JApA...43...82G, 2022JApA...43...11G}. Due to the bright moon phase during the burst observation, we only detected GRB 230204B in the $r$ and $i$-band images from the first night. For the remaining filters, we could only establish upper limits. Photometric calibration is performed using the standard stars from the Pan-STARRS catalog.

\subsubsection{{\it Swift}/UVOT}

The \swift/UVOT began settled observations of the field of \thisgrb 80.6\,ks after the MAXI trigger \citep{2023GCN.33265....1S}. No optical afterglow was detected by UVOT. Photometry was obtained from co-added UVOT image mode data. The source counts were extracted from the UVOT stacked images using a source region of 5 arcsec radius. Background counts were extracted using an aperture of 20 arcsec in radius located in a source-free region near the GRB. We use the {\it Swift} tool \sw{uvotsource} to obtain the background-subtracted count rates using source and background apertures of 5 and 20 arcseconds in radius, respectively. They were converted to magnitude using the UVOT photometric zero points \citep{bre11, poole}.

\subsubsection{Radio Emission}

We observed GRB 230204B on 8 February 2023 (3.6 days post-burst), with the Australia Telescope Compact Array (ATCA, Project code CX527) at a wide range of frequencies centered on  5.5, 9.0, 16.7, 21.2, 33, and 35\,GHz, each with a 2048 MHz-wide band. We reduced the visibility data using standard routines in \texttt{MIRIAD} \citep{1995ASPC...77..433S}. We used a combination of manual and automatic RFI flagging before calibration, conducted with \texttt{MIRIAD} tasks \mbox{} \texttt{uvflag} and \texttt{pgflag}, respectively. We used PKS~1934$-$63 to determine the bandpass response for frequency bands 5.5 and 9 GHz, and PKS~0727$-$115 as the bandpass calibrator for 16.7, 21.2, 33, and 35\,GHz bands. We used PKS~1934$-$63 to calibrate the flux density scale for all frequency bands. We used PKS~1256$-$220 to calibrate the time-variable complex gains for all epochs and frequency bands. After calibration, we inverted the visibilities using a robust weighting of 0.5 and then used the CLEAN algorithm \citep{1980A&A....89..377C} on the target source field using standard \texttt{MIRIAD} tasks \texttt{INVERT}, \texttt{CLEAN}, and \texttt{RESTOR} to obtain the final images. 

We detect the radio counterpart at 16.7 and 21.2 GHz at a position consistent with the BOOTES/MASTER/GIT optical counterpart position \citep{2023GCN.33269....1S}. For all other frequencies (5.5, 9.0, 33.0, 35.0, 0.842, and 0.943 GHz), we report 5-sigma upper limits (see Table \ref{tab:radio} of the appendix). The GRB position was also observed with the Australian SKA Pathfinder by the Rapid ASKAP Continuum Survey \citep[RACS;][]{2020PASA...37...48M} at 0.842, and 0.943 GHz on 2023 July 1 and 2024 Jan 5. We find no radio detections in preliminary data and report 5-sigma upper limits of ~1.5 mJy at 0.842 and 0.943 GHz. There were also no archival radio detection pre-burst within 1 arcminute in surveys: the National Radio Astronomy Observatory VLA Sky Survey \citep[][]{1998AJ....115.1693C} at 1.4 GHz, the Sydney University Molonglo Sky Survey \citep[][]{2003MNRAS.342.1117M} at 843 MHz, the Very Large Array Sky Survey \citep[][]{2020PASP..132c5001L} at 3 GHz or the RACS survey. We measured a VLASS 5-sigma upper limit of 0.69 mJy at 3\,GHz.

\begin{figure}[!ht]
\centering
\includegraphics[scale=0.36]{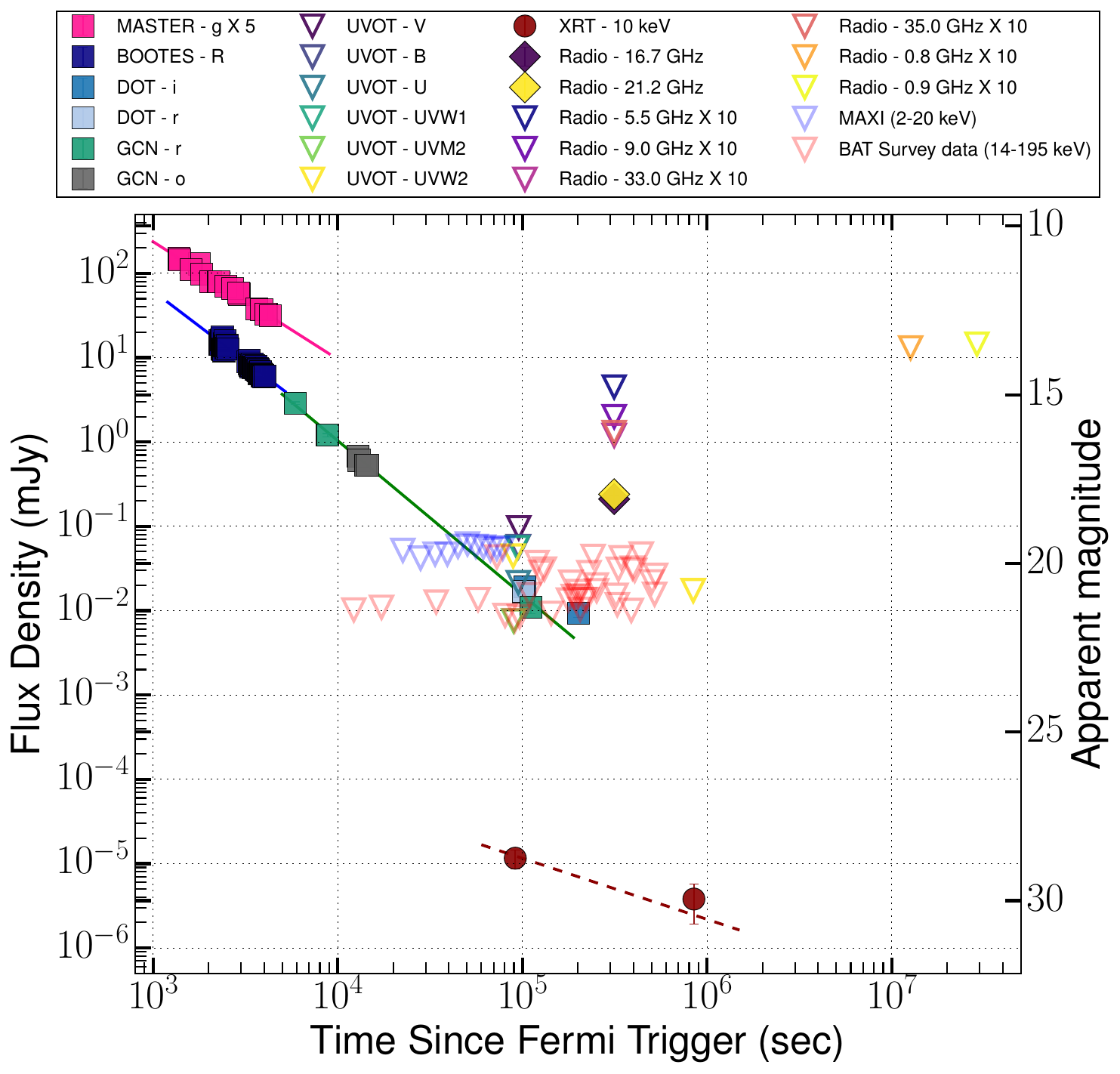} 
\caption{Broad-band composite afterglow flux density light curve of \thisgrb. The solid lines are the best-fit power-law function to optical data (temporal decay indices of -1.67 $\pm$ 0.15 in $R$ and -1.83 $\pm$ 0.05 in $r$ bands, respectively) and the dashed line shows the best-fit power-law function to X-ray data (temporal decay index of -0.72 $\pm$ 0.63). The upper limits for X-ray afterglow search using MAXI (2-20 keV) and BAT Survey data (14-195 keV) are shown using light blue and red color markers, respectively.}
\label{fig:afterglowlc}
\end{figure}

\subsubsection{{\it Swift} BAT Survey Data}

\swift/XRT did not observe the location of X-ray afterglow of \thisgrb until 91.5\,ks post-burst. Therefore, to search for the hard X-ray early afterglow, we used the \texttt{BatAnalysis} python package \citep{BatAnalysis} to analyze \swift BAT survey data from 2023-02-03 to 2023-02-11. We analyzed survey data, where the location of the GRB had at least a partial coding fraction of $\sim 19\%$ on the BAT detector plane. As expected, the GRB afterglow was not detected in the BAT survey data, which is typical since hard X-ray afterglows have never been detected at such late times. However, we were able to place 5$\sigma$ upper limits on the emission in the 14-195 keV energy band. In Figure \ref{fig:afterglowlc}, we show the flux upper limits derived from individual BAT survey snapshots, which correspond to shorter time bins and thus shallower limits. To obtain more stringent constraints, we also performed a mosaiced analysis using 1-day bins for the two days after the trigger, yielding deeper $5\sigma$ upper limits of $\lesssim 2 \times 10^{-9}$ erg sec$^{-1}$ cm$^{-2}$. Although the BAT survey provides the deepest available constraints in the 14–195 keV range during this period, the obtained limits are not sufficiently deep to impose meaningful physical constraints on the afterglow. In particular, the derived $5\sigma$ upper limits are significantly above the flux level expected from a simple extrapolation of the contemporaneous \swift/XRT spectrum to hard X-rays. Thus, while these upper limits confirm the absence of an unusually bright hard X-ray component, they are too shallow to provide additional constraints on the standard afterglow emission.

Based on the above observations, we have shown the broadband composite afterglow light curve of \thisgrb in Figure \ref{fig:afterglowlc}.

\section{Results} 
\label{sec:results}

\subsection{Prompt emission temporal and spectral characterization}

\begin{figure*}
\centering
\includegraphics[scale=0.45]{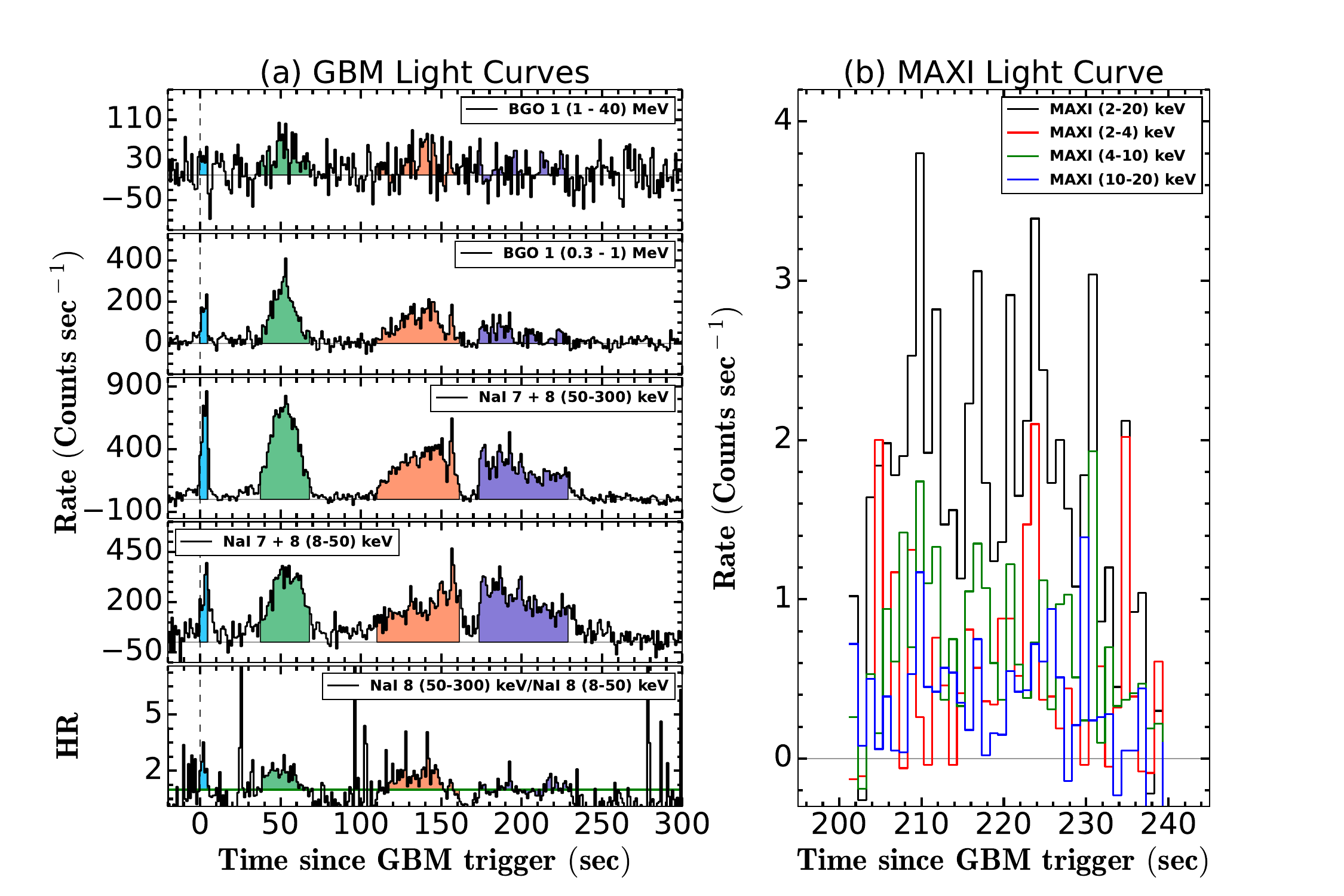} 
\caption{(a) The prompt emission background-subtracted light curve of \thisgrb, obtained using \fermi GBM across different energy channels: NaI 7+8 (8-30 keV), NaI 7+8 (50-300 keV), BGO 1 (0.3-1 MeV), and BGO 1 (1-40 MeV). The shaded regions represent the time intervals used for time-integrated spectral analysis for each of the four emission episodes. The hardness ratio (HR) evolution is shown as a function of time. (b) The MAXI light curve in 2-20 keV, 2-4 keV, 4-10 keV, and 10-20 keV energy bands, illustrating that MAXI observed only a later emission of \thisgrb.}
\label{fig:promptlc}
\end{figure*}

For the analysis of \fermi GBM data, we utilized the \fermi GBM Data Tools\footnote{\url{https://fermi.gsfc.nasa.gov/ssc/data/analysis/gbm/gbm_data_tools/gdt-docs/}} to extract temporal information. We selected specific NaI (NaI 7 and NaI 8) and BGO (BGO 1) detectors for the analysis based on their optimal viewing angles relative to the GRB's position to maximize the signal-to-noise ratio. For the NaI detectors, we focused on two energy bands: 8-50 keV and 50-300 keV, allowing us to explore both low and high-energy photon populations. The 8-50 keV band was chosen to provide a more comprehensive coverage of the soft X-ray emission, while the 50-300 keV band captures the higher-energy component. For BGO 1, two bands were used: 0.3-1 MeV and 1-40 MeV, to investigate higher-energy photons. The prompt emission background-subtracted light curve of \thisgrb, obtained using \fermi GBM across different energy channels, reveals a complex multi-pulse structure, as shown in Figure \ref{fig:promptlc}. This GRB exhibits four distinct emission episodes, each separated by periods of quiescence, with the shaded regions highlighting the intervals selected for time-integrated spectral analysis. The time windows (pulse 1: \fermiT - 0.80 to \fermiT+ 4.97 sec, pulse 2: \fermiT+ 37.00 to \fermiT+ 68.15 sec, pulse 3: \fermiT+ 109.45 to \fermiT+ 161.90 sec, and pulse 4: \fermiT+ 172.58 to \fermiT+ 229.69 sec) for these intervals were determined using the Bayesian Block algorithm, which adaptively identifies significant temporal structures in the light curve based on statistical criteria. The energy-dependent light curves show that the temporal structure and intensity vary significantly across different energy bands, reinforcing the diverse nature of GRB light curves. Notably, the hardness ratio (HR) demonstrates evolution over time, indicating spectral changes during the burst. The inset plot shows the MAXI light curve, covering the 2-20 keV energy range, further subdivided into 2-4 keV, 4-10 keV, and 10-20 keV. Although MAXI detected only part of the GRB, its observations provide valuable insights into the lower-energy emission and suggest a much softer nature of the last pulse of \thisgrb. 

We analyzed the time-integrated (\fermiT-0.80 to \fermiT+229.69 sec) \fermi spectrum of \thisgrb using the Multi-Mission Maximum Likelihood framework (3ML) to evaluate its spectral characteristics. We explore several phenomenological models, including the \sw{Band} function, \sw{Cutoff power-law} (CPL), and combinations with a \sw{Blackbody} component, which is theoretically expected to originate from the photospheric emission of GRBs \citep{2010ApJ...709L.172R, 2013ApJ...765..103L}. For each model, we employed Bayesian spectral fitting to obtain the Deviance Information Criterion (DIC) values, allowing us to systematically compare the model fits. The results indicate that the \sw{Band} function provides a statistically better fit to the time-integrated spectrum compared to the \sw{CPL} alone, as indicated by the DIC values. Furthermore, the addition of a \sw{Blackbody} component alongside the \sw{Band} function yielded an even lower DIC value, suggesting that this model combination better describes the observed spectral features.

In addition to the time-integrated spectrum, we analyzed the spectral evolution across four distinct emission episodes of \thisgrb, fitting each episode with the same set of phenomenological models. Our findings show that the first and third episodes are best described by the \sw{CPL} + \sw{Blackbody} model, while the second and fourth episodes favor the \sw{Band} + \sw{Blackbody} combination based on DIC values. The DIC values and the best-fit spectral parameters for both the time-integrated analysis and individual episodes are summarized in Table \ref{DIC} and Table \ref{bestfitparameters}. These results highlight the diversity in spectral shapes across the burst duration and provide valuable insights into the complex emission mechanisms underlying \thisgrb, supporting the presence of a thermal component in combination with non-thermal models.

\begin{table*}
\centering
\caption{The DIC values for both the time-integrated analysis and individual episodes. The boldface shows the best-fit model.}
\label{DIC}
\begin{tabular}{l|cccc}\hline				
\textbf{Time-intervals}& & &\textbf{DIC value} &\\
(from \fermiT in sec)&\textbf{Band}&\textbf{CPL}&\textbf{BB+Band}&\textbf{BB+CPL}\\  
\hline 
-0.80-229.69& 7118.17& 7123.59& {\bf 7094.00} & 7103.16\\
-0.80-4.97& 3245.26& 3245.68&3220.72 & {\bf 3214.46} \\
37.00-68.15&5083.02 &5091.67 & {\bf5058.11} & 5077.71\\
109.45-161.90& 5655.64 & 5660.37 & 5620.64 & {\bf 5606.26}\\
172.58-229.69& 5601.18 & 5606.86 & 5577.01 & {\bf 5575.45} \\
\hline
\end{tabular}
\end{table*}

\begin{table*}
\centering
\caption{The best-fit spectral parameters for both the time-integrated analysis and individual episodes.}
\label{bestfitparameters}
\begin{tabular}{l|cccccc}\hline				
\textbf{Time-intervals}& & & &\textbf{Best-fit parameters} & &\\
(from \fermiT in sec)&\textbf{$\it \alpha_{\rm pt}$}&\textbf{\Ep (\keV)}&\textbf{$\it \beta_{\rm pt}$}&\textbf{$\it \Gamma_{\rm CPL}$}& \boldmath $E_{\rm c}$ (\keV)& {kT (keV)}\\  
\hline 
-0.80-229.69& -0.78$^{+0.03}_{-0.07}$ & 519.44$^{+81.71}_{-19.58}$ & -3.37$^{+0.16}_{-3.58}$ & -- & --&15.08$^{+29.28}_{\rm unconstrained}$\\
-0.80-4.97& --& --&-- & -0.38$^{+0.05}_{-0.11}$ & 259.10$^{+56.09}_{-13.35}$ & 24.98$^{+24.90}_{-1.05}$\\
37.00-68.15&-0.65$^{+0.03}_{-0.05}$ & 661.98$^{+75.23}_{-21.83}$ & -2.61$^{+0.18}_{-0.24}$&--&-- &15.51$^{+30.46}_{\rm unconstrained}$\\
109.450-161.90& --& --& --&-0.70$^{+0.01}_{-0.07}$ & 514.31$^{+111.17}_{-16.22}$& 14.01$^{+46.29}_{-0.85}$\\
172.58-229.69&-- &-- & --& -0.80$^{+0.01}_{-0.09}$& 239.09$^{+58.10}_{-2.48}$ &10.62$^{+32.45}_{\rm unconstrained}$\\

\hline
\end{tabular}
\end{table*}

GRBs are known for their significant spectral evolution and high variability, with emission properties often varying notably across individual pulses within multi-pulse bursts. \cite{Liang2021} conduct a comprehensive time-resolved Bayesian spectral analysis of well-defined GRB pulses observed by \fermi. They observe a trend toward spectral softening over time, largely reflected in the gradual decrease of the low-energy power-law index. We compared the time-resolved spectral analysis of \thisgrb with the sample studied by \cite{Liang2021}. We also observe that the low-energy power-law indices of \thisgrb become softer over time (hard to soft), with a distinct spectrum evolution from the first to later episodes (see Figure \ref{fig:trscomparison}). Such a spectral evolution feature of prompt emission could be explained as photospheric emission is most commonly observed near the trigger time, while synchrotron-like emission becomes more prevalent in later times, indicating the coexistence of multiple emission mechanisms as the burst evolves \citep{Liang2021}.

\begin{figure}[ht]
\centering
\includegraphics[scale=0.365]{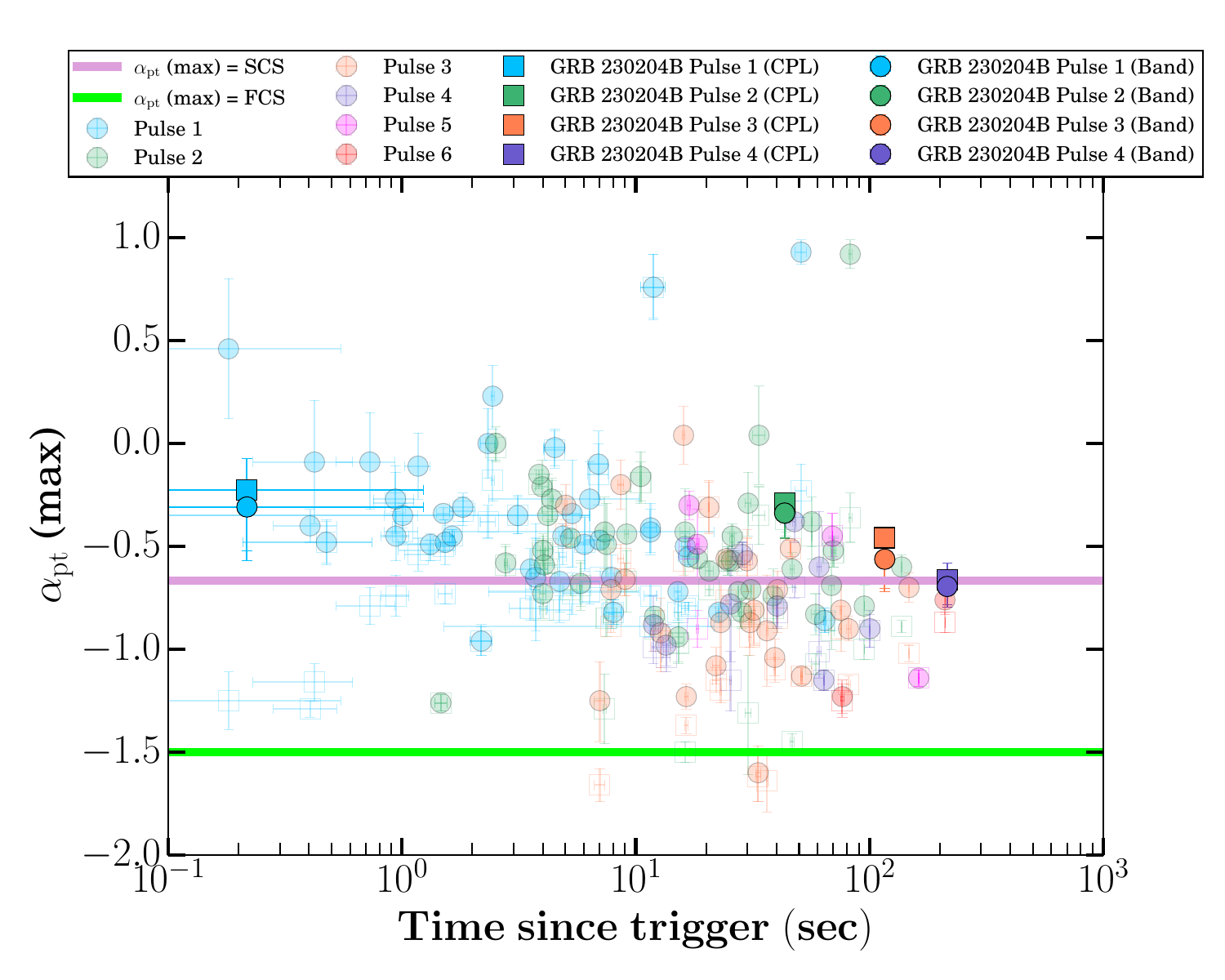} 
\includegraphics[scale=0.36]{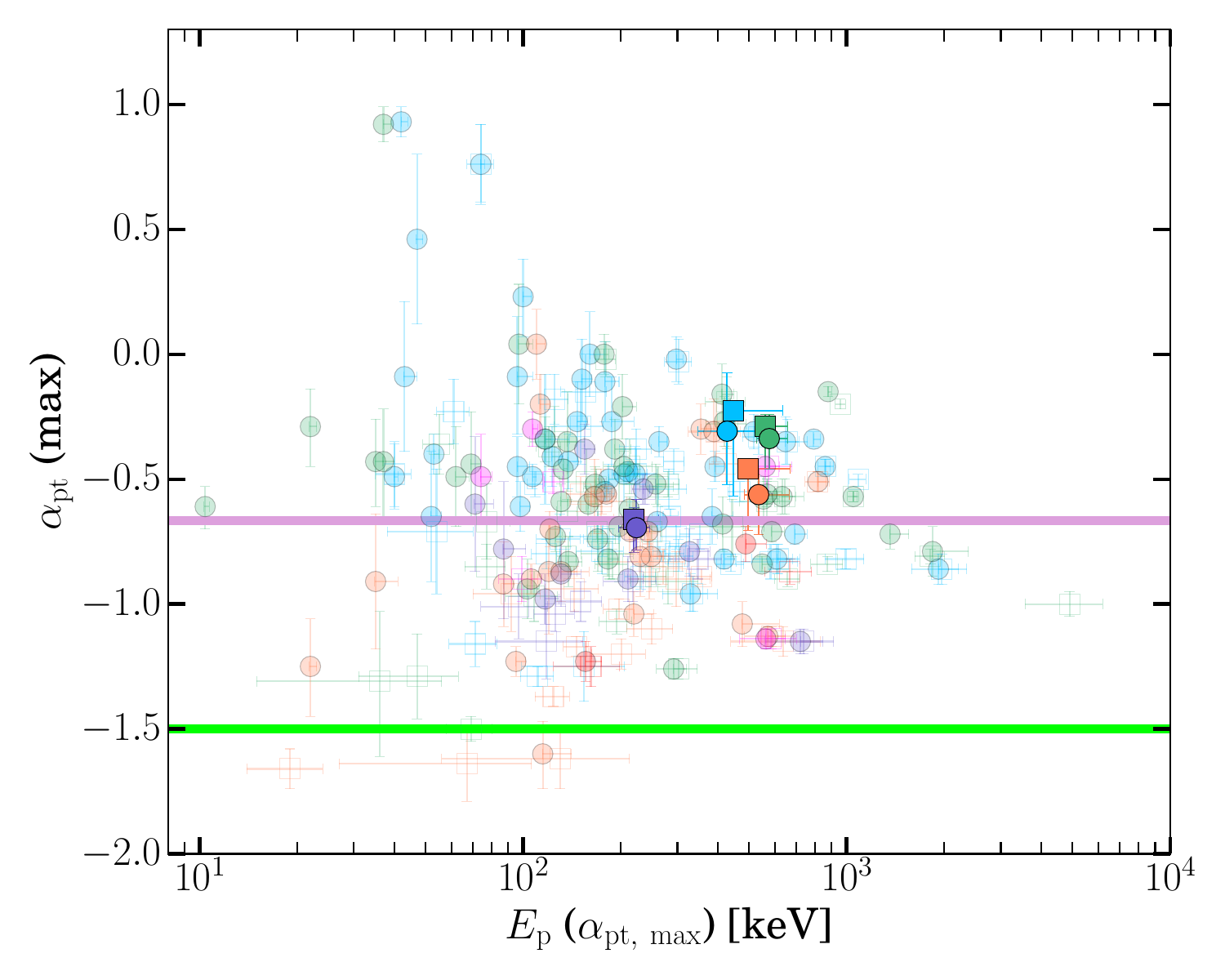} 
\caption{Spectral evolution of \thisgrb during prompt emission detected by \fermi GBM. The top panel shows the evolution of the maximum value of low-energy power-law indices during each pulse of \thisgrb as a function of trigger time. The pink and lime-colored horizontal lines show the synchrotron line of death and the fast-cooling synchrotron emission line, respectively. The bottom panel shows the evolution of the maximum value of low-energy power-law indices during each pulse of \thisgrb as a function of peak energy corresponding to the maximum value of $\alpha$. The data points for spectral parameters of individual pulses of other multi-pulsed \fermi GRBs obtained from \cite{Liang2021}, are also shown.}
\label{fig:trscomparison}
\end{figure}

\subsection{Pulse-wise prompt emission correlation and analysis}
\label{amati}

The pulse-wise analysis of GRB prompt emission provides a valuable perspective on the Amati and Yonetoku correlations (key relationships used to explore the physical properties of GRBs), enhancing our understanding of GRB spectral evolution and energetics on a finer timescale \citep{2020ApJ...898...42C}. By examining each pulse independently, rather than averaging over the entire prompt emission phase, we gain insights into the distinct characteristics of individual emission episodes within a single burst. In the Amati correlation, which relates the isotropic energy ($E_{\rm \gamma, iso}$) to the spectral peak energy in the rest frame \citep{2006MNRAS.372..233A}, each pulse may fall along or diverge from the established long and short GRB trends, providing information about the pulse-specific energy release mechanisms. Such an analysis not only clarifies the evolutionary trajectory of the GRB emission but also tests the applicability of the Amati correlation at sub-burst levels \citep{2013MNRAS.436.3082B}.

In the top-left panel of Figure \ref{fig:promptcorrelation}, we illustrate the positioning of each pulse (plotted as colored squares) of \thisgrb within the established Amati correlation. For reference, samples of long and short bursts from \cite{2020MNRAS.492.1919M} are represented by orange and blue circles, respectively. The linear fits corresponding to these groups are indicated by solid orange and blue lines, with 3$\sigma$ uncertainties depicted by shaded bands around each fit. The time-integrated ($E_{\rm \gamma, iso}$ = 1.92 $\times 10^{54}$ erg) shows that \thisgrb follows the Amati plane of long GRB populations. However, when considering the episodes individually, we find that while later pulse Episode 4 remains broadly consistent with the long GRB population, Episodes 1, 2, and 3 fall into an intermediate region between the long and short GRB distributions, indicating a less well-defined classification.

The top-right panel of Figure \ref{fig:promptcorrelation} presents \thisgrb positions within the Yonetoku correlation framework, a relation between peak energy and the isotropic peak luminosity ($L_{\rm \gamma, iso}$) in the rest frame \citep{2004ApJ...609..935Y}. Here, the short and long GRB samples from \cite{2012MNRAS.421.1256N} are shown by blue and orange circles, with the 3$\sigma$ scatter around the population's best-fit line visualized by a shaded area. Similar to the Amati plane, the overall burst lies on the upper side of the Yonetoku correlation for long GRBs.

\subsubsection{Spectral Peak-Duration Distribution Analysis} 

Long-duration GRBs are generally characterized by a softer spectrum relative to the typically harder spectrum observed in the majority of short-duration GRBs, as demonstrated in previous studies (e.g., \citealt{1993ApJ...413L.101K}). To analyze this characteristic for \thisgrb, we measured both the time-integrated and episode-wise spectral peak energy (an indicator of spectral hardness) alongside the \tninty duration (the time during which 90\% of the burst's total fluence is detected, \tninty values are 214.02 sec, 4.86 sec, 24.32 sec, 41.98 sec, and 51.46 sec for time-integrated duration, episode 1, episode 2, episode 3, and episode 4, respectively). We then plotted the pulse-wise values on the \Ep-\tninty plane to compare with a larger dataset of GRBs detected by the \fermi GBM, as illustrated in the bottom panel of Figure \ref{fig:promptcorrelation}.

Our results show that the time-integrated properties of \thisgrb exhibit a relatively soft \Ep value and duration longer than 2~sec, placing the burst clearly within the Type II (collapsar) GRB population. However, the episode-wise analysis reveals interesting diversity in spectral and temporal properties: Episode 1 lies an intermediate region in both the \Ep-\tninty and Amati correlation (see Section \ref{amati}) plane, with relatively short duration (\tninty $\sim$ 4.86 sec), while Episodes 2 and 3 show progressive evolution toward longer durations and softer spectra. Episode 4 firmly places itself within the long GRB distribution with the softest spectrum ($\alpha$ $\sim$ -0.80) and longest duration (T$_{90}$ $\sim$ 51.46 sec). This diversity likely reflects evolving emission mechanisms throughout the burst. The spectral evolution from photosphere-dominated to synchrotron-dominated emission supports hybrid jet models where thermal and non-thermal components coexist \citep{Liang2021}. Alternatively, for collapsar with residual stellar envelopes, the jet's interaction with material at different radii can produce distinct emission episodes with varying durations and spectral properties \citep{2003ApJ...584..390W, 2003ApJ...586..356Z}.

Our analysis demonstrates that individual emission episodes within a single long GRB can exhibit durations and spectral properties overlapping with the short GRB population and provides evidence for complex, evolving emission physics throughout the burst duration. This finding adds to growing evidence that the traditional 2-second duration threshold for GRB classification may be overly simplistic and that progenitor identification requires consideration of additional 
properties beyond \tninty alone \citep{2025arXiv250800142G}. GRB 230204B exemplifies how such multi-layered approaches reveal complex emission physics that would remain hidden in time-integrated analyses alone. This underscores the importance of pulse-resolved studies for understanding the diversity of GRB emission mechanisms and jet physics.

\subsection{Afterglow analysis and correlation}
\label{afterglowanalyis}

Based on the multi-wavelength follow-up observations campaign ($\sim$ 1.3\,ks to 335 days post burst) of \thisgrb, we noted that BOOTES robotic and DOT observations have a well-sampled light curve in optical $R$ and $r$ filters, respectively. We fitted flux density optical $R$ and r-bands light curves using a simple power-law function with a temporal decay index of -1.67 $\pm$ 0.15 ($\chi^2$/DOF = 20.17/47) and -1.83 $\pm$ 0.05 ($\chi^2$/DOF = 8.25/2), respectively. Furthermore, we combined all the optical data (MASTER, BOOTES, DOT, and a few points from GCN) to get a more well-sampled light curve, and it could also be fitted with a simple power-law function with a temporal decay index of -1.84 $\pm$ 0.02 ($\chi^2$/DOF = 81.13/58). We noted that the optical temporal index is steeper than the typically observed decay slope of $\sim$ -1 \citep{2022ApJ...940..169D}. On the other hand, the X-ray afterglow light curve of \thisgrb has very limited (the last XRT data point has a marginal detection with 3.53$\sigma$) data points as \swift followed the burst during the late phase and could be fitted with a temporal decay index of -0.72 $\pm$ 0.63 ($\chi^2$/DOF = 0.58/1).  

\begin{figure}[ht]
\centering
\includegraphics[scale=0.35]{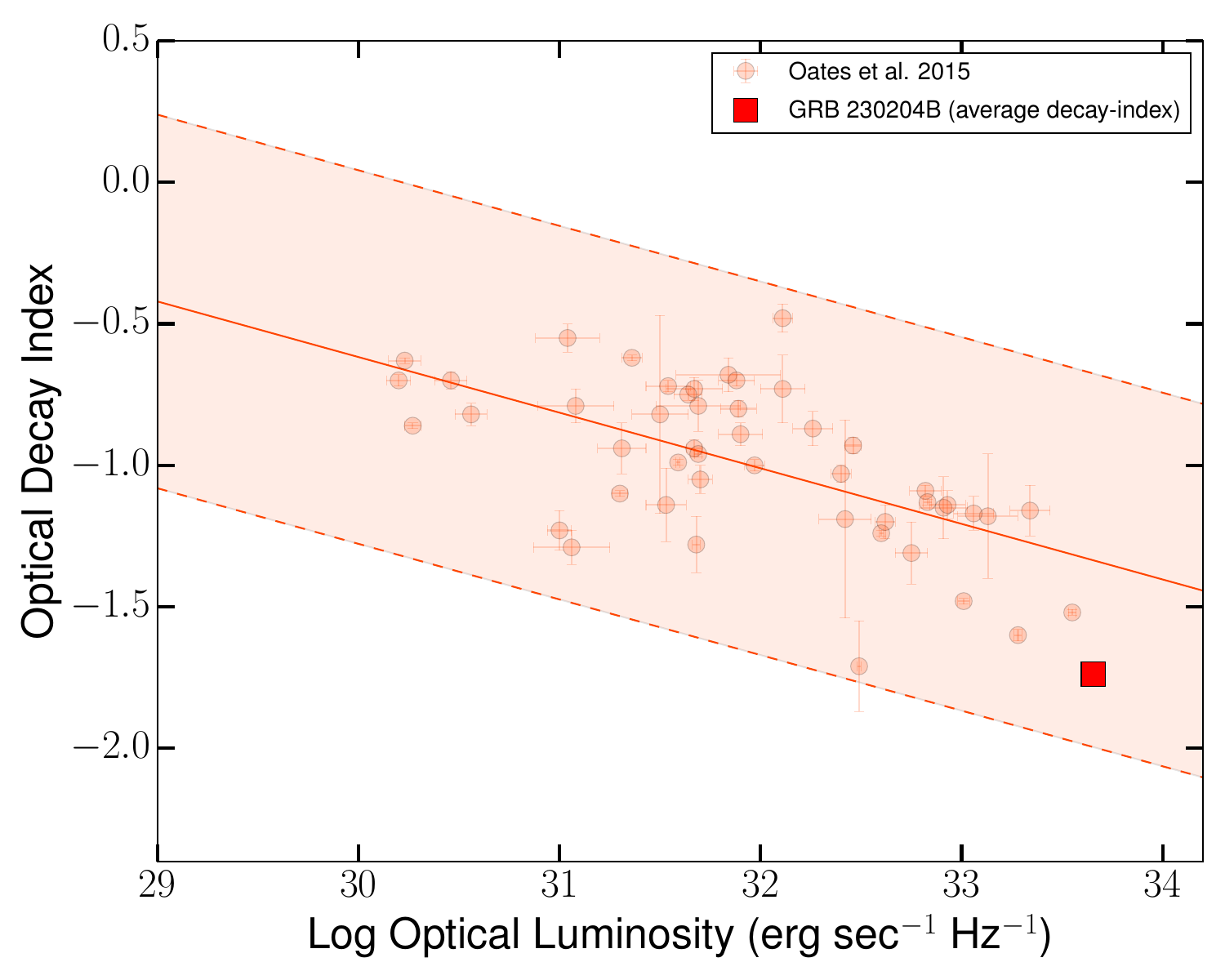} 
\caption{The correlation between the average optical decay index and optical luminosity (at 200-sec post burst) for a sample of GRBs \cite{2012MNRAS.426L..86O, 2015MNRAS.453.4121O}, with GRB 230204B highlighted in red. This plot illustrates the diverse afterglow properties of GRBs, where the decay index reflects the temporal evolution of the optical afterglow and the luminosity indicates the energy output. GRB 230204B’s position in the parameter space aligns with the general trend observed in long-duration GRBs, showcasing very high optical luminosity and a steep decay rate. The solid indicates the best-fit regression analysis for the correlation in this sample, and the shaded region represents the 3$\times$ of root mean square deviation.}
\label{fig:afterglowlccorrelation}
\end{figure}

The correlation between the average optical decay index and optical luminosity provides crucial insights into the physical mechanisms governing GRB afterglows and their surrounding environments. \cite{2012MNRAS.426L..86O, 2015MNRAS.453.4121O} examined the possible correlation between the average optical temporal decay indices and optical luminosity (at 200-sec post-burst) for GRBs with optical afterglows observed by \swift/UVOT. They found evidence of anti-correlation between these two parameters (200-sec post-burst) as shown in Figure \ref{fig:afterglowlccorrelation}. This correlation indicates that optical luminous bursts decay more rapidly than optical faint bursts. We measured optical luminosity and temporal decay index for \thisgrb and overlaid it onto the luminosity-decay correlation. GRB 230204B exhibits a relatively steep decay rate ($\sim$ 1.74 average decay index) and high optical luminosity (at the first epoch of afterglow detection). We noted \thisgrb, highlighted in red in the figure, aligns well with the global correlation observed for long-duration GRBs and lies towards the edge (see Figure \ref{fig:afterglowlccorrelation}). Possible physical explanations of this correlation are discussed in section \ref{brightnesscomparison}.

\subsection{Broadband afterglow modeling of \thisgrb}
\label{model}

\begin{table*}[!ht]
\centering
\caption{Physical parameters of the afterglow for GRB 230204B, derived through Bayesian fitting using the \texttt{Redback} and \texttt{afterglowpy} frameworks.}
\label{afterglowpy}
\begin{tabular}{lcccc}\hline				
\textbf{Model}& Prior &\textbf{Top hat}&\textbf{Gaussian}&\textbf{Gaussian Core}\\
\hline 
$\theta_{\mathrm{v}}~(\mathrm{rad})$&[0, $\pi$/2] & 0.003$^{+0.002}_{-0.001}$ & 0.004$^{+0.002}_{-0.001}$ & 0.005$^{+0.001}_{-0.001}$ \\
$\theta_{\mathrm{core}}~(\mathrm{rad})$&[0.0001, 0.1] & 0.006$^{+0.003}_{-0.002}$ &  0.006$^{+0.002}_{-0.001}$ & 0.007$^{+0.001}_{-0.001}$ \\
$\theta_{\mathrm{w}}~(\mathrm{rad})$& [1, 8] &-- & 4.020$^{+2.676}_{-2.131}$ & 4.714$^{+1.567}_{-1.766}$ \\
$\log_{10}~E_{0} ({\mathrm{erg}})$&[52, 57] & 55.621$^{+0.229}_{-0.262}$& 55.529$^{+0.159}_{-0.187}$ & 55.143$^{+0.113}_{-0.150}$ \\
$p$& [2.0, 3.5] & 2.280$^{+0.020}_{-0.022}$ & 2.267$^{+0.018}_{-0.017}$  & 2.252$^{+0.017}_{-0.017}$ \\
$\log_{10}~n_{\mathrm{ism}} ({\mathrm{cm}}^{-3}$)& [-3, 4] & 1.449$^{+1.036}_{-1.002}$& 1.987$^{+0.823}_{-0.696}$ & 2.685$^{+0.450}_{-0.447}$  \\
$\log_{10}~\epsilon_{B}$ & [-7, 0] & -5.669$^{+0.650}_{-0.651}$ & -5.899$^{+0.500}_{-0.520}$ &  -6.133$^{+0.397}_{-0.318}$\\
$\log_{10}~\epsilon_{e}$ & [-5, 0] &-0.525$^{+0.229}_{-0.209}$ & -0.353$^{+0.170}_{-0.149}$ &  -0.198$^{+0.102}_{-0.107}$\\
$\xi_{N}$&& 1.0 & 1.0 & 1.0\\
ln $Z$& &-101.31$\pm$0.08& -107.50$\pm$0.08 &-110.53$\pm$0.08\\
ln BF& &7638.16 & 7631.97 & 7628.94 \\
\hline
\end{tabular}
\end{table*} 

The broadband afterglow modeling involved fitting the observed data across multiple wavelengths, including radio, optical, and X-ray bands. Such broadband modeling helps to understand the physical parameters of the jet, such as its opening angle, energy, and the surrounding medium's density. The afterglow emission is generally explained by the synchrotron radiation model \citep{1998ApJ...497L..17S, mes97, sari99}. In this model, relativistic electrons are accelerated by the GRB shockwave spiral in the magnetic field, emitting radiation across the electromagnetic spectrum. The synchrotron spectrum is characterized by several break frequencies (self-absorption, peak, and cooling frequencies), which evolve with time and affect the observed light curves at different wavelengths.

For the afterglow modeling of \thisgrb, we performed fitting using \sw{Redback} software \citep{2024MNRAS.531.1203S} via importing afterglow jet models from \sw{afterglowpy}\footnote{It currently does not account for certain features, including reverse shock emission, an external wind medium (n $\propto$ r$^{-2}$), or synchrotron self-absorption.} \citep{2020ApJ...896..166R} to explore various jet structure models, including the Tophat, Gaussian, and GaussianCore models. The Bayesian inference framework provided by \sw{Redback} allowed us to test different priors and achieve robust parameter estimation. Specifically, we employed the \sw{PyMultiNest} sampler to explore the parameter space and used well-defined priors for key parameters such as the observer angle ($\theta_{v}$), core angle ($\theta_{c}$), initial energy (E$_{0}$), and interstellar medium density (n$_{ISM}$). Due to the limited number of data points, we have fixed the bulk Lorentz factor ($\Gamma$ $\sim$ 800), constrained using the $E_{\rm \gamma, iso}$-Lorentz factor correlation \citep{2010ApJ...725.2209L}. The priors used in the analysis include uniform distributions for the energy, core angle, ambient medium density, and a sine prior for the observer's viewing angle.

\begin{figure*}
\centering
\includegraphics[scale=0.4]{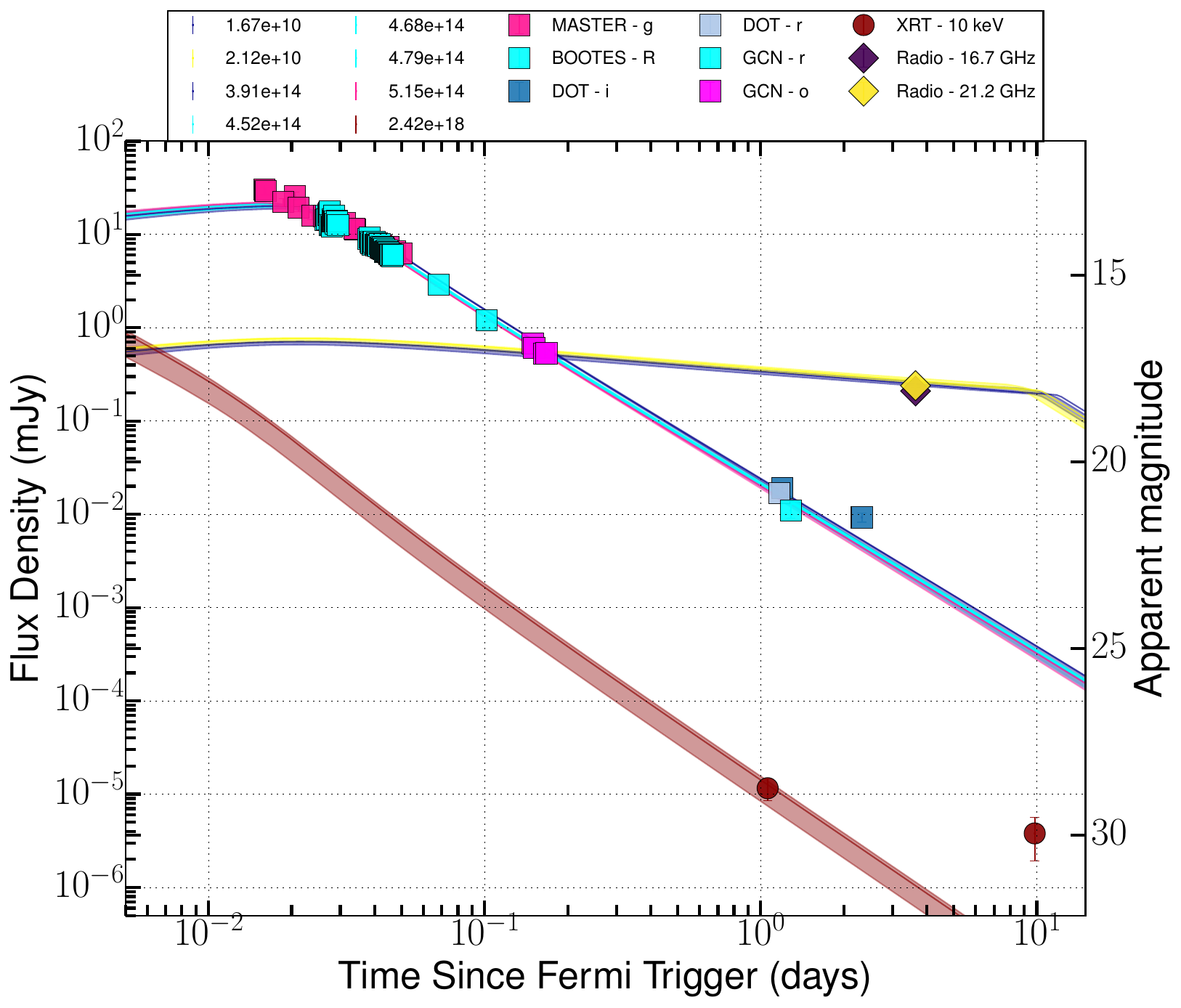} 
\caption{Broadband composite afterglow light curve of \thisgrb, encompassing data from radio, optical, and X-ray observations. The light curve is modeled using a simple top-hat forward shock jet model in a homogeneous interstellar medium (ISM). The best-fit model, shown as solid lines, provides a good representation of the observed multi-wavelength data. The last XRT data point (a marginal detection with 3.53$\sigma$), observed $\sim$ 10 days post-burst, shows a deviation from the model prediction.}
\label{fig:afterglowmodel}
\end{figure*}

The Top-Hat model provided the best fit to the data (see Figure \ref{fig:afterglowmodel}), with log evidence (ln $Z$) of -101.31 $\pm$ 0.08 and a Bayes factor (ln BF) of 7638.16 as indicated by the Bayesian evidence comparison. These results strongly suggest that the jet structure for \thisgrb is best described by a simple Top-Hat profile (see Table \ref{afterglowpy}). The posterior distributions (see corner plot as shown in Figure \ref{fig:afterglowmodelcorner} of the appendix) constrained viewing angle of $\theta_{v}$ 0.003 rad and a core angle $\theta_{c}$ 0.006 rad (limit as no jet break is observed), suggesting a narrow, highly collimated jet structure. The inferred initial energy is relatively high at $\log_{10}~(E_{0})$ $\sim$ 55.62 erg, and the ISM density is constrained to $\log_{10}~(n_{\mathrm{ism}})$ $\sim$ 1.45 cm$^{-3}$. We also modeled the afterglow of GRB 230204B with \sw{VegasAfterglow} \citep{2025arXiv250710829W}, assuming a top-hat jet in an ISM-like medium, and obtained parameters consistent with those derived using \texttt{afterglowpy}. The best-fit values are: $\theta_{\mathrm{v}} = 0.004^{+0.001}_{-0.001}$ rad, $\theta_{\mathrm{core}} = 0.014^{+0.002}_{-0.002}$ rad, $\log_{10}(E_{0}) = 55.01^{+0.15}_{-0.12}$ erg, $p = 2.30^{+0.02}_{-0.02}$, $\log_{10}(n_{\mathrm{ISM}}) = 2.05^{+0.14}_{-0.16}$ ${\mathrm{cm}}^{-3}$, $\log_{10}\,\epsilon_{B} = -5.45^{+0.25}_{-0.23}$, and $\log_{10}\,\epsilon_{e} = -0.89^{+0.07}_{-0.08}$.

The Top-Hat jet model provides a robust description of \thisgrb's afterglow, indicating a narrow, highly collimated jet. This is consistent with the high fluence observed in the early afterglow phase. The tightly constrained parameters for $\theta_{v}$ and $\theta_{c}$ indicate that we are likely observing the jet from a viewing angle close to the core, providing a relatively straightforward interpretation in the context of synchrotron emission from a top-hat jet model. The last XRT data point at $\sim$10 days post-burst shows a deviation from the afterglow model prediction. This observation corresponds to a marginal detection (3.53$\sigma$) with a count rate of $\sim 0.001$ ct sec$^{-1}$ (exposure of 4533.2 sec), close to the sensitivity limit of \swift-XRT. Due to the limited photon statistics, we could not reliably constrain the spectral parameters. Considering the excess flux is intrinsic, it could arise from additional physical processes such as late-time energy injection \citep{2006ApJ...647.1238J}, refreshed shocks \citep{2022MNRAS.513.2777K}, or wavelength-dependent emission components. However, in the absence of further X-ray or multiwavelength coverage at comparable epochs, the true origin of this excess remains uncertain.

A notable feature in the fit is the very low value of $\epsilon_{B}$, the magnetic field fraction, which was constrained to $\log_{10}~\epsilon_{B}$ = -5.67$^{+0.65}_{-0.65}$. Such low magnetic fields can possibly arise from shock compression of a weakly magnetized ambient medium \citep{2009MNRAS.400L..75K, 2015ApJS..219....9W}, and similar values have been observed in GRBs such as GRB 190530A \citep{Gupta1}, GRB 171205A \citep{2021MNRAS.503.1847L}, GRB 130427A \citep{2016ApJ...818..190F} and a few others \citep{2009MNRAS.400L..75K, 2010MNRAS.409..226K, 2014ApJ...785...29S}, highlighting the diversity in GRB environments. For GRBs, when $\epsilon_{e}$ $> >$ $\epsilon_{B}$, there might be Synchrotron Self-Compton (SSC) contribution to afterglow \citep{2021MNRAS.505.1718J}. However, in the case of GRB 230204B, our afterglow modeling using \sw{Redback} and \sw{VegasAfterglow} (see below) suggests that the SSC effects appear negligible, consistent with findings for similar GRBs \citep{Gupta1, 2024ApJ...969..146W}. This is further supported by the temporal decay slopes, and no significant GeV emission (consistent with synchrotron-dominated regimes). However, we cannot completely rule out some SSC contribution, particularly at early radio epochs, where current limited data may lack the sensitivity to detect weak SSC components. Future studies focusing on a broader sample of bright GRBs with low $\epsilon_{B}$, combined with well-sampled broadband light curves (from gamma-ray to radio) and afterglow modeling that includes SSC contributions, have the potential to shed light on their physical origins and diverse emission mechanisms.

Furthermore, to investigate alternative scenarios, we modeled the afterglow of GRB 230204B using the \sw{VegasAfterglow} framework \citep{2025arXiv250710829W}, a high-performance and versatile code that self-consistently treats forward and reverse shock dynamics, synchrotron and inverse Compton radiation, and arbitrary circumburst density profiles. Unlike commonly used tools such as \sw{afterglowpy}, \sw{VegasAfterglow} supports wind-like media, allowing us to probe environments shaped by massive-star mass loss. We performed Markov Chain Monte Carlo (MCMC) fitting of the broadband afterglow light curves under the assumption of a stellar-wind environment (see Table \ref{afterglowpywind}) with several jet structures (top-hat, power-law, and Gaussian). The best-fitting wind model with a top-hat jet and resulting posterior distributions for key physical parameters are shown in Figures \ref{fig:afterglowmodelwind} and \ref{fig:afterglowmodelcornerwind} of the appendix, respectively. This modeling yields constraints on the isotropic-equivalent energy, jet opening angle, electron distribution index, and other microphysical parameters, indicating a luminous afterglow produced by a narrowly collimated jet.

To discriminate between ISM and wind-like circumburst environments for \thisgrb, we performed a statistical comparison using reduced $\chi^{2}$ values for both the models. For the ISM top-hat model: $\chi^{2}/\mathrm{dof} = 153.51/70 = 2.19$, and for the Wind top-hat model: $\chi^{2}/\mathrm{dof} = 174.37/71 = 2.46$. While both reduced $\chi^{2}$ values are greater than 1, the ISM model provides a notably better fit. To quantify the statistical significance of this improvement, we applied an F-test. The F-statistic yields $F \approx 9.5$ (F-value), and corresponds to a p-value of $\sim 0.003$, indicating that the ISM model is statistically significantly better at the $>99.7\%$ confidence level. This indicates that the improvement is not merely due to random fluctuations but reflects a genuine preference for the ISM scenario.

Beyond the statistical framework, we examined the standard closure relations for model discrimination. Using the electron energy distribution indices obtained from our afterglow modeling: $p_{\rm ISM} = 2.280^{+0.020}_{-0.022}$ (ISM scenario), $p_{\rm wind} = 2.125^{+0.021}_{-0.020}$ (wind scenario), we calculated the predicted temporal and spectral indices. For the slow-cooling relations, the closure relations are:

\begin{align}
\beta(\nu_m < \nu < \nu_c) &= \frac{p-1}{2}, \qquad 
\beta(\nu > \nu_c) = \frac{p}{2}, \\
\alpha_{\rm ISM}(\nu_m < \nu < \nu_c) &= \frac{3(p-1)}{4}, \qquad
\alpha_{\rm ISM}(\nu > \nu_c) = \frac{3p-2}{4}, \\
\alpha_{\rm wind}(\nu_m < \nu < \nu_c) &= \frac{3p-1}{4}, \qquad
\alpha_{\rm wind}(\nu > \nu_c) = \frac{3p-2}{4}.
\end{align}

For the ISM-derived $p$, these relations predict $\beta_{\rm ISM} \simeq 0.640^{+0.010}_{-0.011}$ or $1.140^{+0.010}_{-0.011}$ and temporal slopes $\alpha_{\rm ISM} \simeq 0.960^{+0.015}_{-0.016}$ or $1.210^{+0.015}_{-0.016}$ for the ($\nu_m < \nu < \nu_c$) and ($\nu > \nu_c$) spectral regimes, respectively. For the wind-derived $p$, we obtain $\beta_{\rm wind} \simeq 0.562^{+0.010}_{-0.010}$ or $1.062^{+0.010}_{-0.010}$, and temporal indices $\alpha_{\rm wind} \simeq 1.344^{+0.016}_{-0.015}$ or $1.094^{+0.016}_{-0.015}$ for the analogous spectral regimes. Notably, both ISM and wind predictions systematically underestimate the observed steep optical decay rate. Consequently, the standard closure relations fail to discriminate the environments, primarily due to the unavailability of well-sampled multiwavelength data. In contrast, our comprehensive afterglow modeling framework provides a more robust and physically complete determination that the ISM scenario is preferred.

\section{Discussion} 
\label{sec:discussion}

\subsection{Spectral evolution and Possible Radiation Mechanism}

The spectral evolution of the prompt emission reveals complex temporal and spectral variability, which provides crucial insights into the nature of the emission mechanisms and the physical conditions in the GRB jet \citep{2014A&A...568A..45B}. Typically, GRB prompt spectra evolve over time, with the peak energy often shifting to lower energies as the burst progresses. This trend, known as ``hard-to-soft" evolution, is commonly observed and is thought to reflect cooling processes within the jet \citep{1986ApJ...301..213N, 1994ApJ...426..604B}. Alternatively, ``tracking" behavior, where \Ep follows the intensity variations of the burst, has also been observed in some cases \citep{1999ApJ...512..693R}, suggesting a possible relationship between the central engine activity and the observed emission \citep{2021MNRAS.505.4086G}. 

The radiation mechanisms driving GRBs are fundamentally linked to the spectral index observed during the prompt emission phase and the overall radiative efficiency. The spectral index $\alpha_{\rm pt}$, which describes the shape of the GRB's spectrum, provides crucial insights into the underlying physical processes. A harder low-energy spectral index (i.e., closer to or exceeding -2/3) typically suggests a dominant photospheric or thermal component, where radiation emerges directly from the photosphere of the relativistic outflow. In contrast, a softer (less than -2/3) often points to non-thermal processes, such as synchrotron emission, where electrons are accelerated in magnetic fields, producing a broad power-law spectrum \citep{2022ApJ...936...12C, 2023MNRAS.519.3201C}.

To investigate the spectral evolution of GRB 230204B, we conducted a time-resolved spectral analysis, selecting time bins based on a Bayesian algorithm for optimal selection. The top panel of Figure \ref{fig:TRS} presents the deviation information criterion (DIC) values, comparing the \sw{Band} function with the CPL for each time-resolved bin. The DIC comparison indicates that the majority of bins—excluding three—are better described by the traditional \sw{Band} model. In the bottom panel of Figure \ref{fig:TRS}, we illustrate the temporal evolution of key spectral parameters, including \Ep, and $\alpha_{\rm pt}$, as derived from the \sw{Band} and \sw{CPL} models for GRB 230204B (see Table \ref{TRS_Table}). These parameters are overlaid on the count-rate light curve of the burst's prompt emission, revealing that the observed \Ep exhibits an intensity-tracking behavior. Harder spectral indices during the first and second episodes might suggest a prominent photospheric component, which often produces harder spectra due to thermal-like emission \citep{2010ApJ...709L.172R}. The softer spectral indices during the later bins of the second, third, and fourth episodes might suggest that a synchrotron-like emission becomes more prevalent in later times, indicating the coexistence of multiple emission mechanisms as the burst evolves \citep{Liang2021, 2018MNRAS.475.1708A}. 

\begin{figure}
\centering
\includegraphics[scale=0.36]{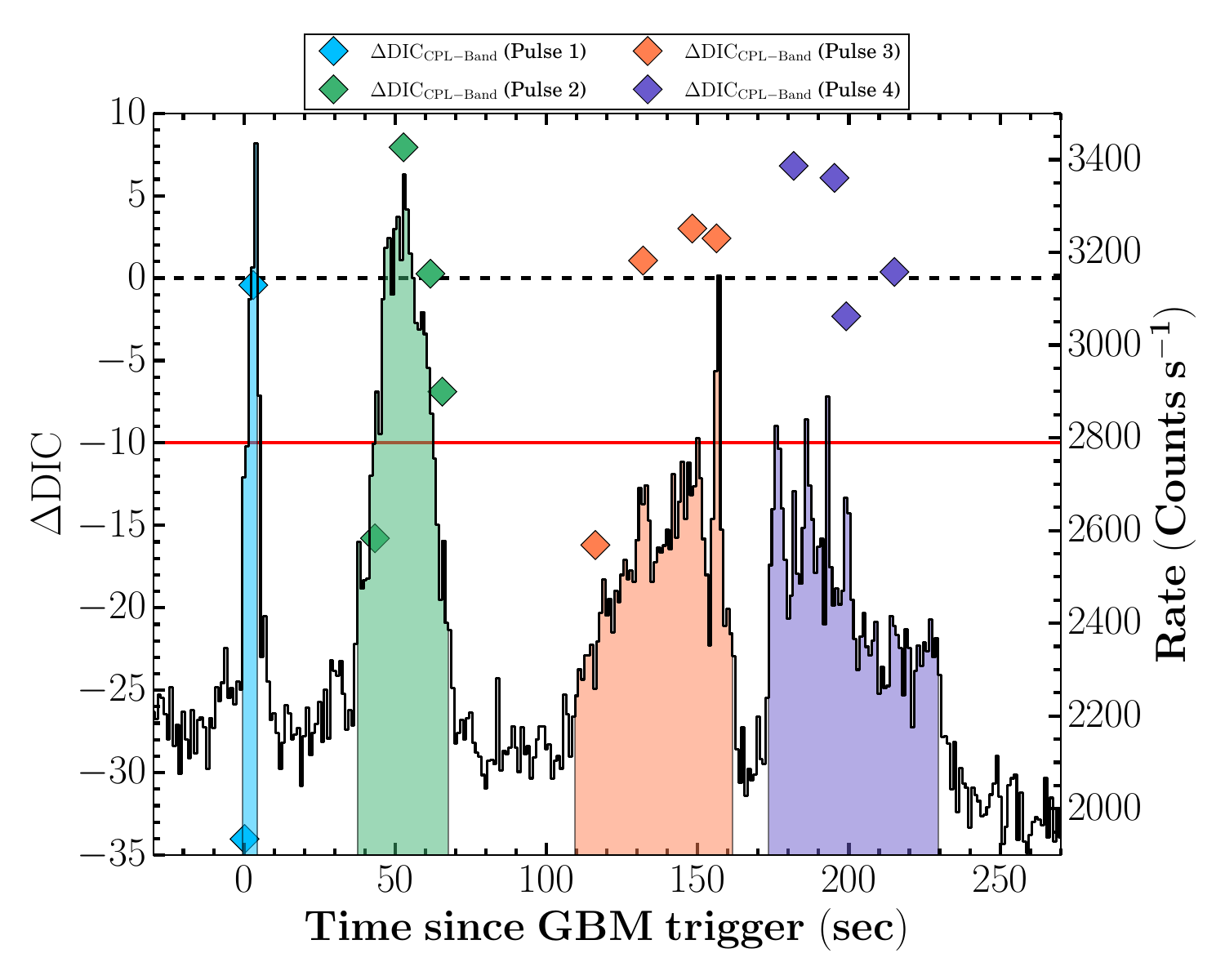} 
\includegraphics[scale=0.36]{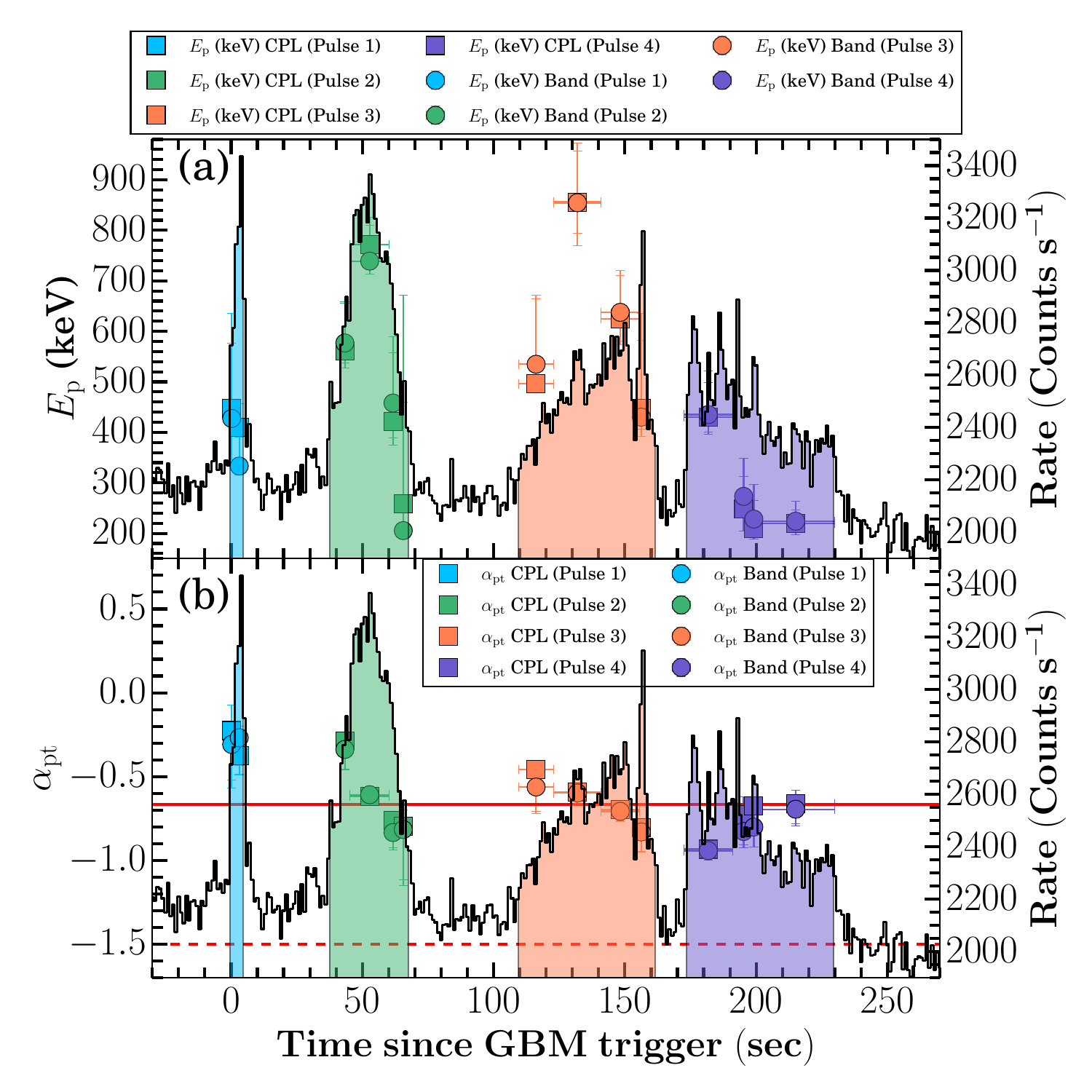} 
\caption{Top panel: The comparison between both phenomenological models using DIC values for individual time bins. Bottom panel: Spectral evolution of \thisgrb during prompt emission detected by \fermi GBM and derived from time-resolved spectral analysis. (a) Variation of the peak energy during each episode of \thisgrb. (b) Changes in the low-energy spectral index, with horizontal dashed and dash-dotted lines representing the synchrotron ``line of death” ($\alpha_{\rm pt}$ = -2/3) and the fast-cooling synchrotron limit ($\alpha_{\rm pt}$ = -3/2), respectively.}
\label{fig:TRS}
\end{figure}

Additionally, we explored correlations among the spectral parameters, examining 1. log(Flux) versus log(\Ep), 2. log(Flux) versus $\alpha_{\rm pt}$, and 3. log(\Ep) versus $\alpha_{\rm pt}$. Pearson correlation analysis was used to quantify the correlation strengths (Pearson correlation coefficient, r) and to assess the statistical significance (p-value) for each parameter pair. We noted a strong correlation for log(Flux) versus log(\Ep) of \sw{Band} and \sw{CPL} models with correlation coefficients equal to 0.69 and 0.73, respectively. We also noted a moderate correlation for log(Flux) versus $\alpha_{\rm pt}$ of \sw{Band} model with a correlation coefficient equal to 0.46, but no correlation for log(\Ep) versus $\alpha_{\rm pt}$.

The radiation efficiency of GRBs is another key factor, reflecting the fraction of the jet's kinetic energy converted into gamma rays. It varies significantly depending on the dominant radiation mechanisms within the jet \citep{2015ApJS..219....9W}. In the photospheric model, radiation is produced as photons escape from the optically thick inner regions of the relativistic outflow. Energy is released at the photosphere, where the flow becomes transparent, leading to thermal emission. Photospheric models tend to predict higher radiation efficiencies, sometimes exceeding 50-60\%, especially if energy dissipation occurs near the photosphere \citep{2009ApJ...700L..47L, 2010MNRAS.407.1033B, 2011ApJ...732...49P, 2013ApJ...765..103L}. In contrast, the synchrotron models (arising from accelerated particles in internal shocks or magnetic reconnection sites) often predict lower efficiencies, typically in the range of a few percent to 30\%, in such cases much of the energy remains in the kinetic form rather than being radiated away \citep{1997ApJ...490...92K, 2000ApJ...537..824S, 2013ApJ...769...69B, 2015PhR...561....1K}. The hybrid model combines thermal photospheric emission with non-thermal synchrotron radiation from shocks or magnetic reconnections within the jet. This model can accommodate both thermal and non-thermal features in GRB spectra. Hybrid models can achieve efficiencies ranging from a few percent to 50\% or higher, as they leverage both thermal and non-thermal emission processes \citep{2011ApJ...726...90Z, 2009ApJ...700L..65Z, 2011ApJ...738...77V}. These discussions suggest that GRBs may involve a complex interplay of thermal and non-thermal processes, with the relative contributions of these mechanisms varying across different bursts and even between different phases of the same burst.

\begin{figure}[ht]
\centering
\includegraphics[scale=0.38]{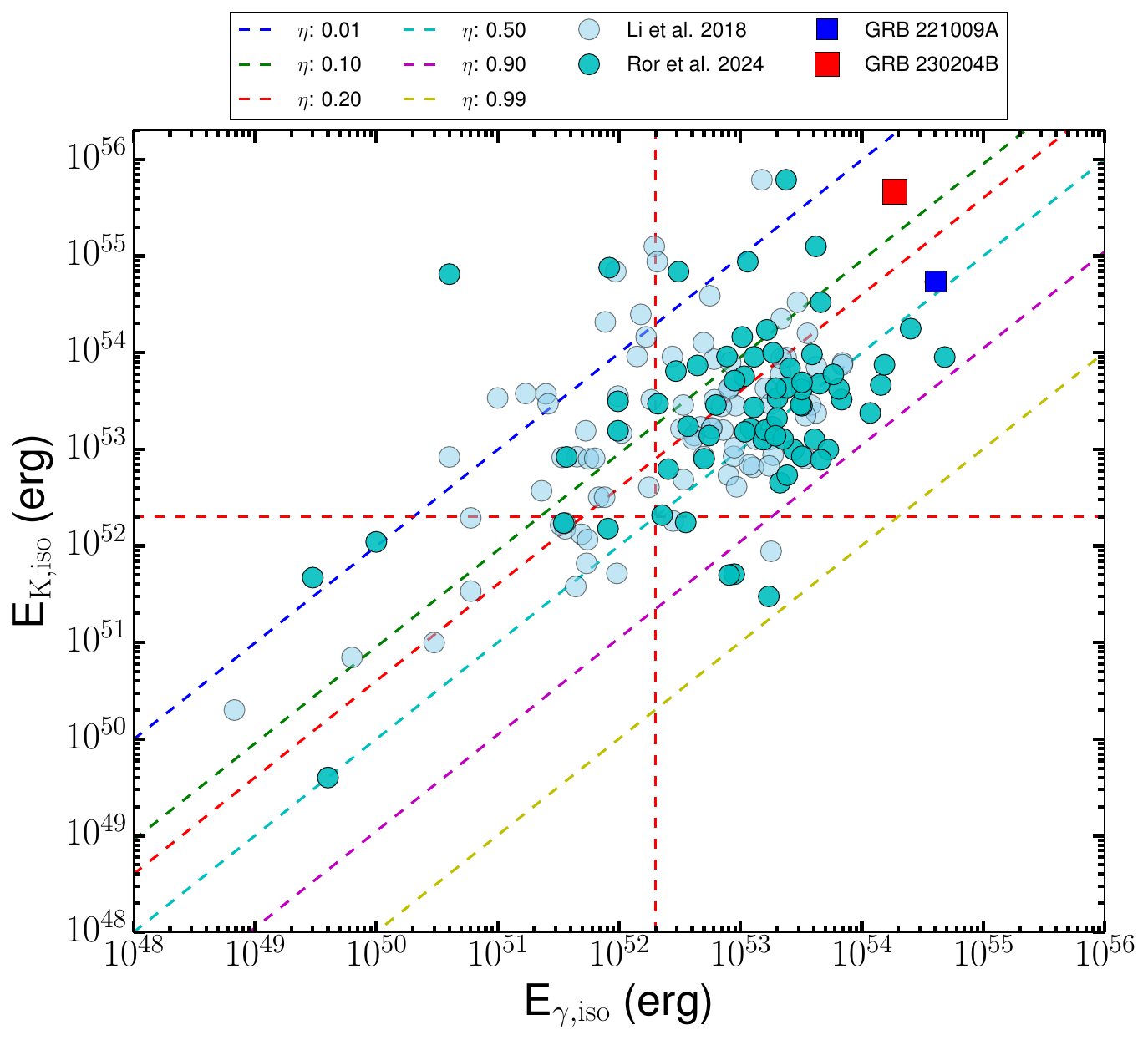} 
\includegraphics[scale=0.38]{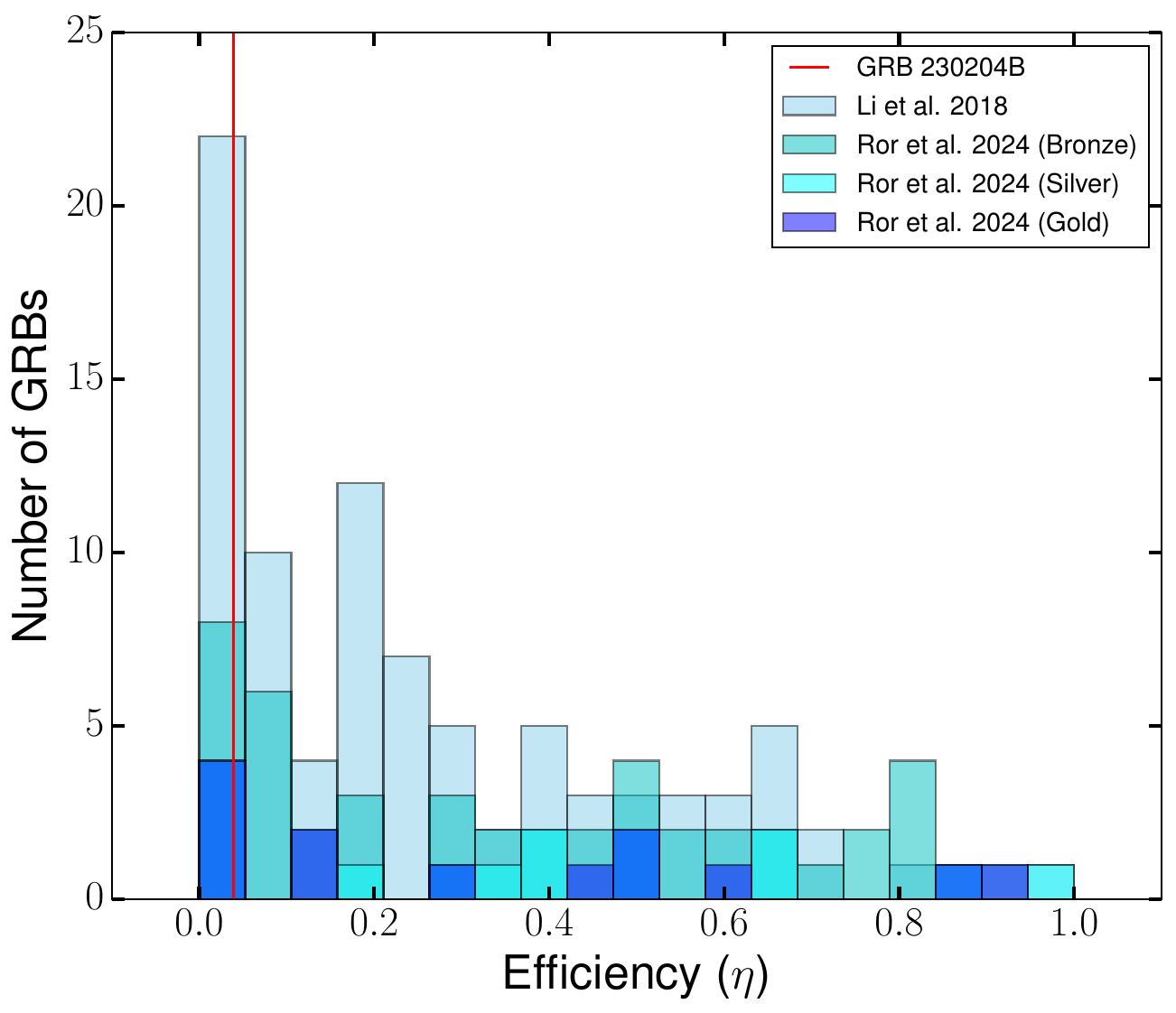} 
\caption{Top: Correlation between the isotropic-equivalent gamma-ray energy ($E_{\rm \gamma, iso}$) and kinetic energy ($E_{\rm K, iso}$, obtained from afterglow modeling) for GRB 230204B, compared to the broader GRB population obtained from \cite{2018ApJS..236...26L, 2024ApJ...971..163R}. GRB 230204B is marked by a red square, indicating its position as a low-efficiency burst. The dashed lines represent the one-to-one correspondence between $E_{\rm \gamma, iso}$ and $E_{\rm K, iso}$. Bottom: Distribution of radiation efficiencies for a sample of GRBs, with GRB 230204B highlighted in red. The figure demonstrates the broad range of efficiencies across the population, with GRB 230204B residing in the low-efficiency tail. The red dashed lines correspond to the magnetar limit.}
\label{fig:efficiency}
\end{figure}

The prompt emission and afterglow observations of GRB 230204B present intriguing insights into its jet structure and radiation mechanisms. Our analysis of the prompt emission reveals a spectral hardening during the initial pulses, with a subsequent softening in later pulses, consistent with a hybrid composition (baryonic + Poynting-flux) of \sw{Band} + \sw{Blackbody} components. This spectral evolution suggests that the jet may consist of both photospheric (thermal) and synchrotron (non-thermal) components, with the harder spectrum at the start dominated by a photospheric component and later phases increasingly influenced by synchrotron radiation \citep{2020MNRAS.493.5218S, 2024ApJ...972..166G}. Additionally, our afterglow modeling indicates a low gamma-ray efficiency of ~4.3\%, suggesting that most of the jet's energy remains in kinetic form with minimal conversion into prompt gamma-ray emission. While the low gamma-ray efficiency supports a hybrid model, where both thermal photospheric and non-thermal synchrotron components coexist, it does not entirely rule out a synchrotron-only scenario \citep{2002MNRAS.332..945R}. Such a model, without any photospheric thermal emission, could also account for the low efficiency, provided the energy dissipation mechanism predominantly drives non-thermal radiation. However, the observed spectral evolution, with the initial hard component transitioning to softer emission, is more naturally explained by the presence of a photospheric contribution early on. Furthermore, the low gamma-ray efficiency aligns well with the characteristics of an ISM-like top-hat jet structure in the afterglow phase. In such a scenario, a top-hat jet interacting with a homogeneous interstellar medium produces a relatively uniform energy distribution across the jet, reducing the likelihood of highly efficient gamma-ray emission \citep{1998ApJ...497L..17S, 2002ApJ...568..820G, 2005ApJ...628..315Z}. This jet structure also favors an afterglow powered predominantly by forward shock emission, which is consistent with the observed light curve and its steady decay pattern.
 
The kinetic energy and $E_{\rm \gamma, iso}$ of GRB 230204B place it within the population of GRBs characterized by relatively low radiation efficiency. The kinetic energy for GRB 230204B, inferred from afterglow modeling, was found to be significantly higher than the isotropic gamma-ray energy, emphasizing that a substantial fraction of the jet's energy remained in kinetic form and was not radiated as gamma rays. We noted that \thisgrb has a higher $E_{\rm K, iso}$ and $E_{\rm \gamma, iso}$ values than expected from magnetar central engine (2 $\times$ 10$^{52}$ erg, \citealt{2011MNRAS.413.2031M}). This supports the rapidly rotating black hole as a possible central engine for this energetic and bright burst \citep{2018ApJS..236...26L, 2024ApJ...971..163R}. The distribution of radiation efficiencies across the GRB population, as shown in Figure \ref{fig:efficiency}, highlights GRB 230204B as the upper edge with a very low efficiency.

\subsection{Brightness Comparison to GRB sample and physical scenarios for \thisgrb}
\label{brightnesscomparison}

\begin{figure*}
\centering
\includegraphics[scale=0.5]{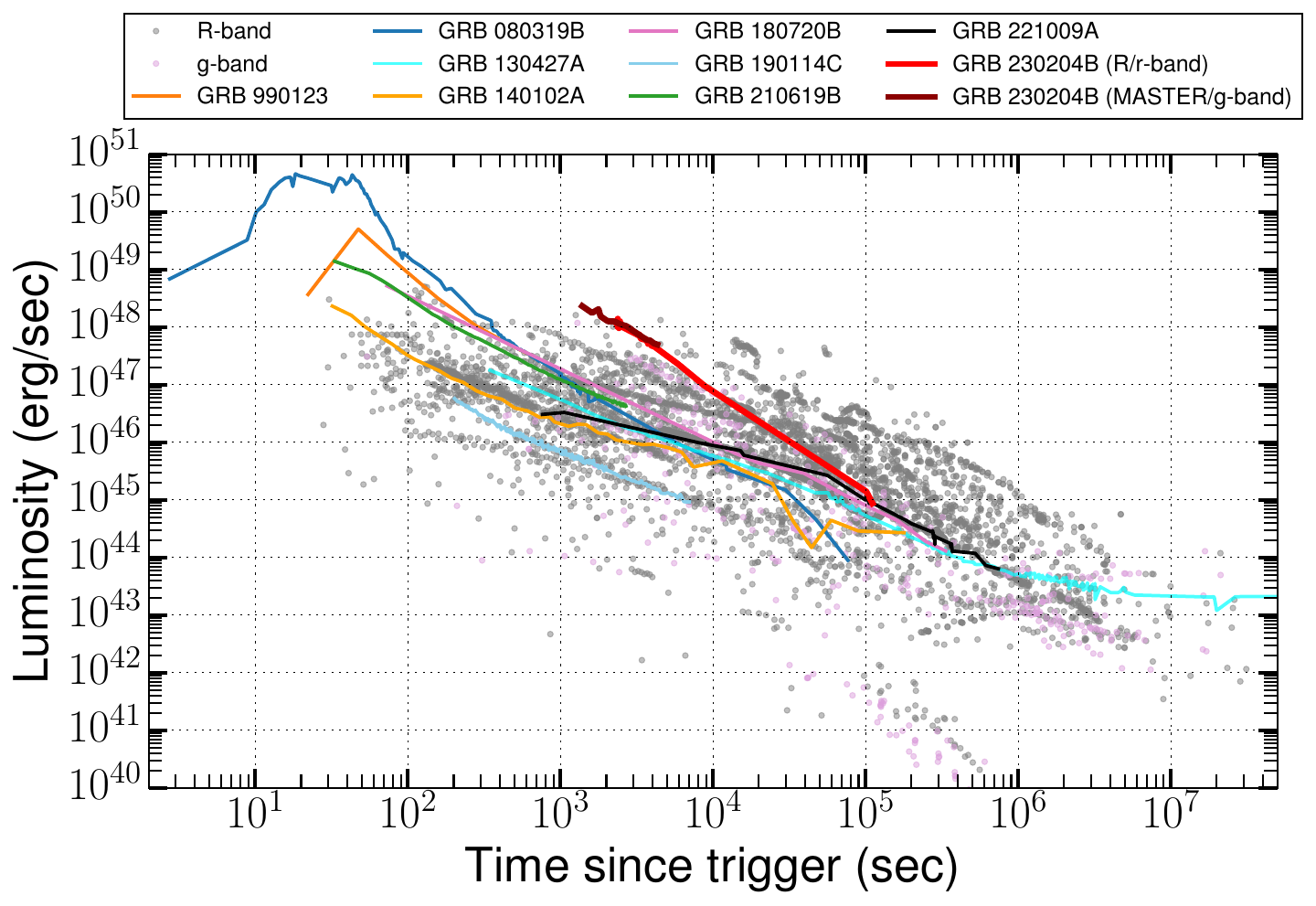} 
\includegraphics[scale=0.5]{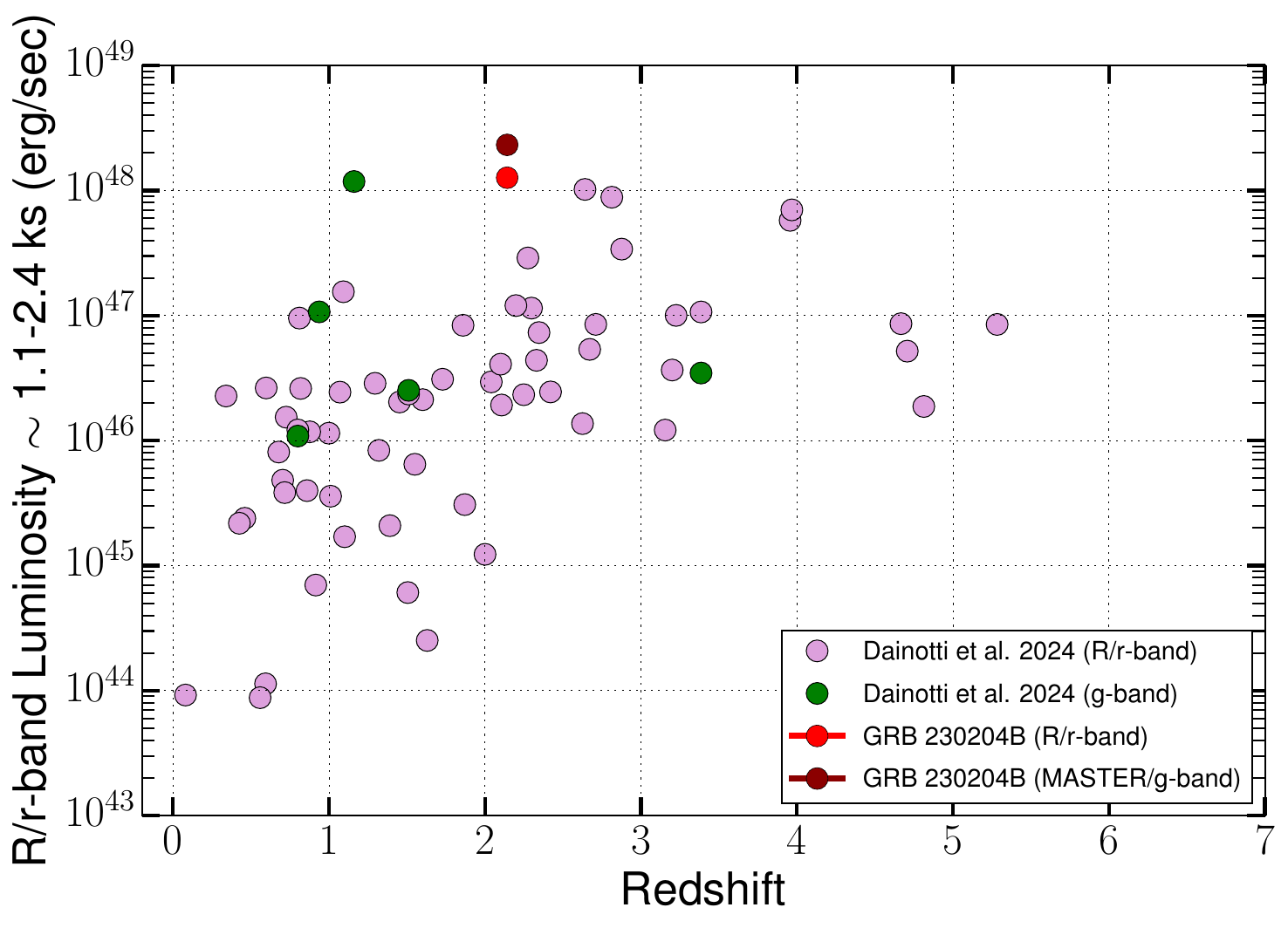} 
\caption{{Top panel:} Temporal evolution of the optical luminosity of GRB 230204B compared with a complete sample of well-studied GRBs with measured redshifts \citep{2024MNRAS.533.4023D}. The red solid curve represents the optical light curve of GRB 230204B, and the points show data from a sample of well-studied GRBs with measured redshifts. The comparison highlights the early extremely luminous optical emission and the subsequent decay phase of GRB 230204B, positioning it among the most luminous optical GRBs detected to date. Specific GRBs, such as GRB 990123, GRB 080319B, GRB 130427A, GRB 140102A, GRB 180720B, GRB 190114C, GRB 210619B, and GRB 221009A, are highlighted with individual lines as they represent notable cases with exceptionally bright afterglows. {Bottom panel:} Distribution of GRBs with measured redshifts as a function of optical luminosity at the epoch of the first optical detection of GRB 230204B. GRB 230204B is marked in red, showcasing its relative optical brightness within the broader GRB population. This figure indicates that GRB 230204B is the most luminous burst at the epoch of its first detection. The time window of 1.1–2.4 ks was chosen for the luminosity comparison to ensure a consistent and well-sampled epoch across all GRBs, corresponding to the period of the first optical detection and robust monitoring of GRB 230204B.}
\label{fig:opticalluminosity}
\end{figure*}

In this Section, we compared the optical, X-ray, and radio afterglow brightness of \thisgrb with a larger population of GRBs with measured redshifts and explored the physical scenarios.

The optical luminosity of \thisgrb (R/r-band and g-band data) has been compared with the population of GRBs with measured redshifts \citep{2024MNRAS.533.4023D}, providing valuable insights into its optical emission characteristics. The temporal evolution of GRB 230204B’s optical luminosity (see Figure \ref{fig:opticalluminosity}) reveals an exceptionally bright initial emission, followed by a decay phase consistent with a forward shock model in a homogeneous interstellar medium (ISM). This places it among the brightest optical GRBs ever observed at the epoch of its detection, making it a standout event in the context of optical afterglow studies. The extreme optical brightness is an order of magnitude higher than other highly energetic bursts ($\sim$ 1.3\,ks post-burst) such as GRB 080319B \citep{2008Natur.455..183R, 2009A&A...504...45P}, GRB 990123 \citep{1999Sci...283.2069C}, GRB 190114C \citep{2021MNRAS.504.5685M, 2021RMxAC..53..113G}, and GRB 221009A. Furthermore, we have also examined the optical luminosity of \thisgrb at the epoch of its detection as a function of redshift for a population of GRBs with measured redshifts \citep{2024MNRAS.533.4023D}. We noted that \thisgrb has the highest optical luminosity ($\sim$ 1.1-2.4\,ks post-burst) with respect to all known GRBs with measured redshifts at the same epoch (see Figure \ref{fig:opticalluminosity}). This suggests that \thisgrb could represent a highly efficient energy transfer to the optical afterglow, or it might be related to the properties of its surrounding environment, jet structure, or progenitor system, contributing to its extraordinary luminosity.

Additionally, we noted that \thisgrb follows an anti-correlation between the optical decay index and optical luminosity \citep{2012MNRAS.426L..86O, 2015MNRAS.453.4121O}. There are several scenarios where such a correlation could be explained. Continued energy injection from a central engine, such as a magnetar or a rapidly spinning black hole, could initially produce a very bright afterglow \citep{2001ApJ...552L..35Z}. However, if the injection is short-lived or if the engine rapidly loses power, the afterglow could exhibit a steep decay as the additional energy source wanes \citep{2001ApJ...552L..35Z, 1998A&A...333L..87D}. Our afterglow analysis did not show evidence of central engine activity (flare, plateau, variability, etc). A reverse shock could also produce a very luminous optical flash \citep{2005ApJ...628..315Z}. If the reverse shock is strong, it can dominate the early afterglow. Once the reverse shock fades, the light curve might steepen significantly, leading to a rapid decay. For \thisgrb, however, our afterglow modeling suggests that a forward shock model could explain the early optical emission. Another possibility involves a structured jet, where a bright core with steep wings might cause a steep decay if the emission transitions from being core-dominated to wing-dominated, or if there is a slight change in the viewing angle \citep{2002MNRAS.332..945R, 2003ApJ...591.1086G}. However, our afterglow modeling suggests that a top-hat jet could explain the afterglow of \thisgrb, so we discarded the structured jet for this burst. If the GRB is in a wind-like medium (density $\sim$ r$^{-2}$), the early afterglow can be very bright \citep{2000ApJ...536..195C}. A change in the density profile (e.g., a transition from a wind to a uniform medium) can also produce a steep decay in the afterglow light curve \citep{2002ApJ...571..779P, 2008ApJ...672..433S}. For GRB 230204B, we modeled the afterglow in both ISM and wind-like environments. While the wind scenario can reproduce some features of the light curve, our broadband analysis shows that the ISM-like medium provides a better overall fit to the data. A jet break occurs when the relativistic beaming angle of the jet becomes wider than the observer's line of sight \citep{2009ApJ...698...43R}. This typically results in a steepening of the afterglow light curve \citep{2008ApJ...675..528L}. An achromatic temporal slope of $\sim$ -2 at late times could indicate that the break occurred; however, for \thisgrb, synchronous broadband data is not available to confirm such behavior, and such an early jet break is not expected, so we discarded the possibility. GRB afterglows depend strongly on the observer's viewing angle relative to the jet's core \citep{1999ApJ...513..679G, 2013ApJ...767..141V}. When an observer views the GRB slightly off-axis but still close to the jet core, the observed afterglow appears very bright initially. This is due to the deceleration of the relativistic jet as it interacts with the surrounding medium, causing the beamed emission to broaden and spread into the observer's line of sight. As the jet slows further and its emission becomes more isotropic, the light curve steepens, producing a rapid decline \citep{2015ApJ...799....3R}. For \thisgrb, our afterglow modeling yields an observer angle of $\theta_{\rm obs} = 0.003^{+0.002}_{-0.001}$ rad and a jet core angle of $\theta_{c} = 0.006^{+0.003}_{-0.002}$ rad. Although the nominal values of the observer and jet core angles suggest that the line of sight is not exactly aligned with the jet axis. However, within the reported uncertainties, the two angles are consistent. This is possibly compatible with the steep early optical decay and high luminosity.

In Figure \ref{fig:radio}, we present a comparative analysis of the X-ray luminosity light curves of GRB 230204B with a sample of other GRBs that have measured redshifts \citep{2024arXiv240904871G}. The X-ray flux data for these GRBs were sourced from the Swift XRT light curve repository\footnote{\url{https://www.swift.ac.uk}}. To enable a direct comparison, the light curves are presented in the rest frame in the k-corrected co-moving bandpass, allowing for a consistent examination of intrinsic luminosity evolution across the sample. GRB 230204B is characterized by limited data, with only two available data points recorded at a late epoch, depicted in red on the plot. Despite the scarcity of data, the observed luminosity for GRB 230204B during this late phase is notably consistent with the average luminosity values observed in the comparative GRB sample at a similar epoch. This suggests that GRB 230204B's X-ray luminosity profile falls within the expected range for GRBs in this phase.

In our analysis, we also examined the radio luminosity of \thisgrb in the context of a broader sample of GRBs with known redshifts, as compiled by Shilling et al. (in preparation). Figure \ref{fig:radio} illustrates the rest-frame 8.5 GHz luminosity light curves for this GRB sample, restricted to measurements with a signal-to-noise ratio (SNR) greater than 2 for clarity. For \thisgrb specifically, we used a 9 GHz observation from ATCA as a close approximation to the 8.5 GHz rest-frame frequency. This 9 GHz upper limit, taken several days post-burst, is plotted alongside the 8.5 GHz light curve distribution. We noted that \thisgrb’s radio upper limit appears on the higher side of the observed luminosity range for GRBs in the sample. The positioning of \thisgrb’s radio data within this distribution provides valuable context for its radio emission profile relative to the larger GRB population. This comparison indicates that \thisgrb’s intrinsic radio luminosity is consistent with the broader sample, suggesting it shares similar radio emission characteristics with typical GRBs.

\section{SUMMARY} 
\label{sec:SUMMARY}

In this study, we conducted a comprehensive analysis of GRB 230204B, a long GRB observed across multiple wavelengths. Our observational campaign utilized both space-based telescopes, including \fermi GBM, MAXI, \swift/BAT, \swift/XRT, and \swift/UVOT, as well as ground-based observatories like BOOTES, MASTER, DOT, and ATCA. The burst displayed a multi-pulse structure (each separated by periods of quiescence) with significant spectral variability, which we explored using time-resolved and time-integrated spectral analyses. Our findings indicate that the \sw{Band} function, when combined with a \sw{Blackbody} component, provided the best fit to the time-integrated spectra (indicating a potential mix of photospheric and synchrotron emission mechanisms), with variations in the preferred models across individual pulses. Furthermore, the time-resolved analysis reveals that the initial pulses of the prompt emission show harder spectra, while subsequent pulses become progressively softer, indicative of a hybrid jet composition.

Furthermore, the afterglow data collected across radio, optical, and X-ray wavelengths showed that GRB 230204B’s luminosity at various stages aligns with typical long-duration GRBs. Optical follow-up using multiple ground-based observatories (mainly using the MASTER and BOOTES robotic telescopes) revealed a rapid, early afterglow decay rate, which was more luminous than most GRBs at a similar epoch. The contributions of the MASTER and BOOTES robotic telescope networks were indispensable for the early detection and characterization of the optical afterglow of GRB 230204B. The scientific impact of the early robotic observations extends beyond simple discovery and can be understood through these key aspects:

The continuous monitoring by the MASTER and BOOTES networks over the first few hours post-burst revealed a smooth power-law decay without evidence of optical flare, reverse shock emission, or other early-time features that have been observed in other bright GRBs (e.g., GRB 080319B \citealt{2008Natur.455..183R}, GRB 990123, \citealt{1999Natur.398..400A}). This forward-shock-dominated evolution, established through early observations, allowing robust constraints on the microphysical parameters, including the circumburst density, jet viewing angle, jet opening angle, and magnetic field properties through broadband modeling (see Section \ref{model}). These parameters are fundamental to understanding the energetics and collimation of this exceptionally luminous GRB, underscoring the importance of robotic networks in the study of transient astronomical phenomena. In particular, without these rapid robotic measurements, the steep optical decay and its placement on luminosity–decay correlations would remain unconstrained. The high optical luminosity and bright afterglow of \thisgrb indicate efficient energy transfer from the burst to the surrounding medium or an environment conducive to high luminosity. Our afterglow modeling suggests that this can be explained within the framework of a top-hat jet.

Broadband afterglow modeling was conducted to investigate the GRB’s jet structure and surrounding environment. Using \sw{afterglowpy}/\sw{Redback}, we evaluated different jet profiles, concluding that a Top-Hat jet model best describes the data, with the inferred viewing and core angles suggesting a narrow, well-collimated jet structure. GRB 230204B exhibits an exceptionally high optical luminosity and an early steep optical decay, consistent with an observer viewing the burst nearly on-axis, as supported by afterglow modeling and a narrow inferred viewing angle. Our afterglow modeling supports a forward shock emission in a homogeneous interstellar medium (ISM), with no evidence of central engine activity. The optical luminosity and temporal decay index were found to align with existing GRB luminosity-decay correlations, further supporting a forward-shock-dominated afterglow. The radiative efficiency derived for GRB 230204B is lower (~4.3\%) compared to estimates reported for typical GRBs. The lower efficiency is primarily due to a lower magnetic field energy fraction, which leads to systematically higher estimates of $E_{\rm K, iso}$. Consequently, this significantly mitigates the long-standing issue of low efficiency associated with internal shock models. We noted that \thisgrb has a higher $E_{\rm K, iso}$ and $E_{\rm \gamma, iso}$ values than expected from magnetar central engine (2 $\times$ 10$^{52}$ erg). This supports the rapidly rotating black hole as a possible central engine for this energetic and bright burst.

In summary, GRB 230204B's low gamma-ray efficiency and hybrid jet composition suggest a radiation mechanism where photospheric and synchrotron processes co-exist within the jet, influencing the spectral evolution of the prompt emission. This case highlights how the efficiency of GRB emissions depends not only on the jet composition and radiation processes but also on the structure of the jet and its interaction with the surrounding environment. Further studies of GRBs with similar spectral evolution and energy efficiency will help clarify the diversity of radiation mechanisms in GRBs and their implications for jet physics and energy dissipation.

Overall, our analysis of GRB 230204B highlights its significance as a benchmark for studying the emission mechanisms, and afterglow evolution of long-duration GRBs. The results not only enrich the broader understanding of GRB diversity but also underscore the indispensable role of broadband, multi-wavelength follow-up observations in probing the complex physical processes governing these events. In particular, this case study demonstrates that the value of robotic telescope networks extends far beyond rapid discovery: their early-time, high-cadence data provide essential constraints on jet dynamics, circumburst environments, and emission mechanisms that would otherwise remain ambiguous. As next-generation and expanded robotic networks become operational, their systematic ability to capture the earliest phases of GRB afterglows will be crucial for refining jet models, improving physical parameter estimates, and advancing our overall understanding of GRB physics and relativistic jet composition. 

\begin{acknowledgments}

We sincerely thank the referee for their insightful and constructive feedback, which has significantly improved the quality of our manuscript. RG was sponsored by the National Aeronautics and Space Administration (NASA) through a contract with ORAU. The views and conclusions contained in this document are those of the authors and should not be interpreted as representing the official policies, either expressed or implied, of the National Aeronautics and Space Administration (NASA) or the U.S. Government. The U.S. Government is authorized to reproduce and distribute reprints for Government purposes notwithstanding any copyright notation herein. RG and SBP acknowledge the financial support of ISRO under AstroSat archival Data utilization program (DS$\_$2B-13013(2)/1/2021-Sec.2). RG, AKR, SBP, and AA acknowledge the BRICS grant DST/ICD/BRICS/Call5/CoNMuTraMO/2023 (G) funded by the DST, India. This research has used data obtained through the HEASARC Online Service, provided by the NASA-GSFC, in support of NASA High Energy Astrophysics Programs. AA also acknowledges the Yushan Young Fellow Program by the Ministry of Education, Taiwan, for financial support. This work made use of data supplied by the UK Swift Science Data Centre at the University of Leicester. This research is based on observations obtained at the 3.6m Devasthal Optical Telescope (DOT), which is a National Facility run and managed by Aryabhatta Research Institute of Observational Sciences (ARIES), an autonomous Institute under the Department of Science and Technology, Government of India. The Australia Telescope Compact Array is part of the Australia Telescope National Facility (\url{https://ror.org/05qajvd42}) which is funded by the Australian Government for operation as a National Facility managed by CSIRO. We acknowledge the Gomeroi people as the Traditional Owners of the Observatory site. Parts of this research were conducted with support from the Australian Research Council Centre of Excellence for Gravitational Wave Discovery (OzGrav), project number CE230100016. AJCT acknowledges support from the Spanish Ministry projects PID2020-118491GB-I00 and PID2023-151905OB-I00 and Junta de Andaluc\'ia grant P20$\_$010168 and from the Severo Ochoa grant CEX2021-001131-S funded by MCIN/AEI/10.13039/501100011033. SS acknowledges support from an STFC PhD studentship, the Faculty of Science and Technology at Lancaster University, and NASA under award number 80GSFC21M0002. MG acknowledges the Academy of Finland project no. 325806. The program of development within Priority-2030 is acknowledged for supporting the research at UrFU (04.89).
\end{acknowledgments}

\clearpage
\facilities{Fermi, MAXI, Swift, MASTER, BOOTES, DOT, ATCA}

\software{astropy \citep{2013A&A...558A..33A, 2018AJ....156..123A},  
          Source Extractor \citep{1996A&AS..117..393B}, GBM-Tool \citep{GbmDataTools}, 3ML \citep{2015arXiv150708343V}, XSPEC \citep{1996ASPC..101...17A}, DAOPHOT-II \citep{1987PASP...99..191S}, IRAF \citep{1986SPIE..627..733T, 1993ASPC...52..173T}, Matplotlib \citep{2007CSE.....9...90H}
          }

\appendix
\restartappendixnumbering

\section{Figures}

\begin{figure}
\centering
\includegraphics[scale=0.35]{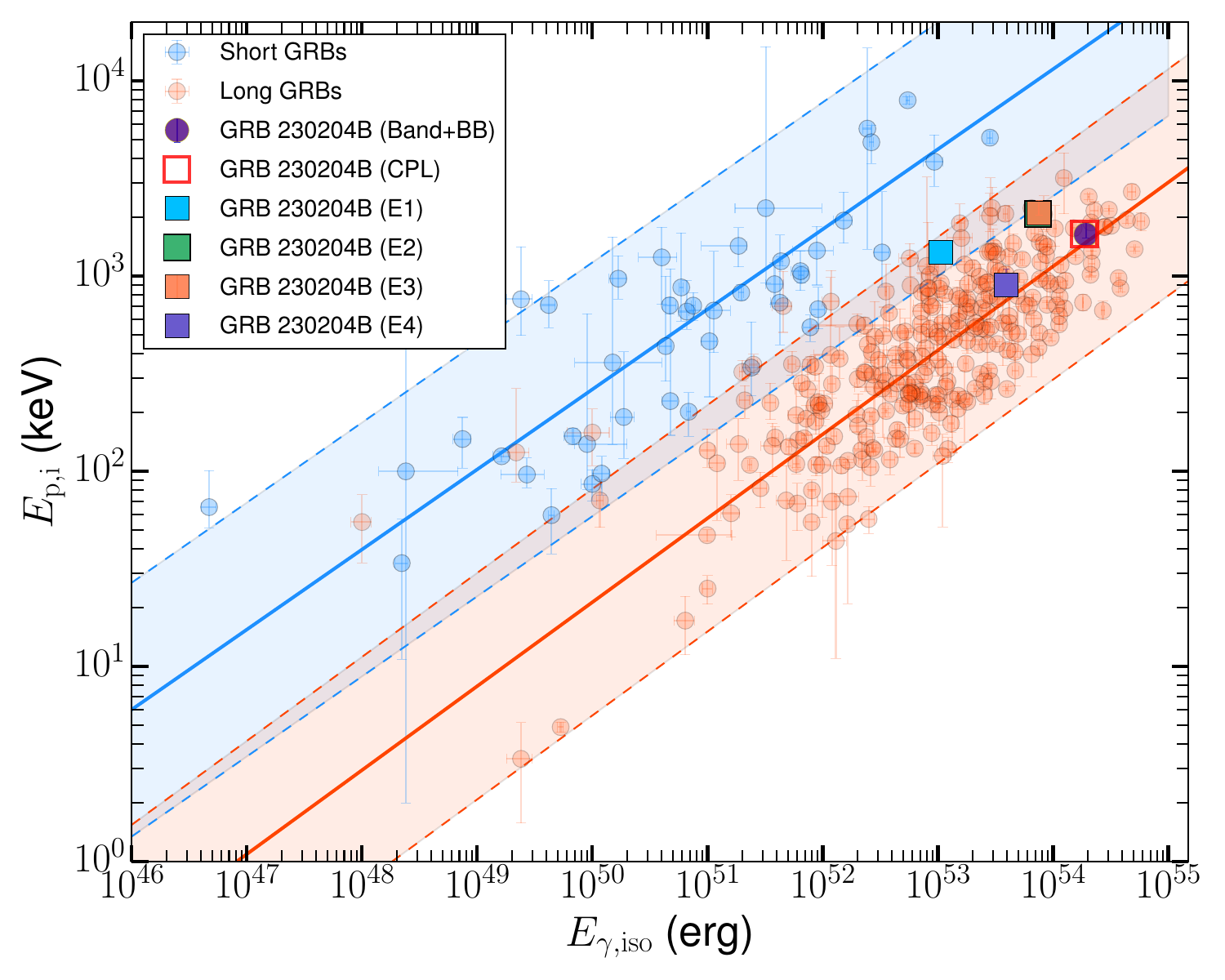} 
\includegraphics[scale=0.35]{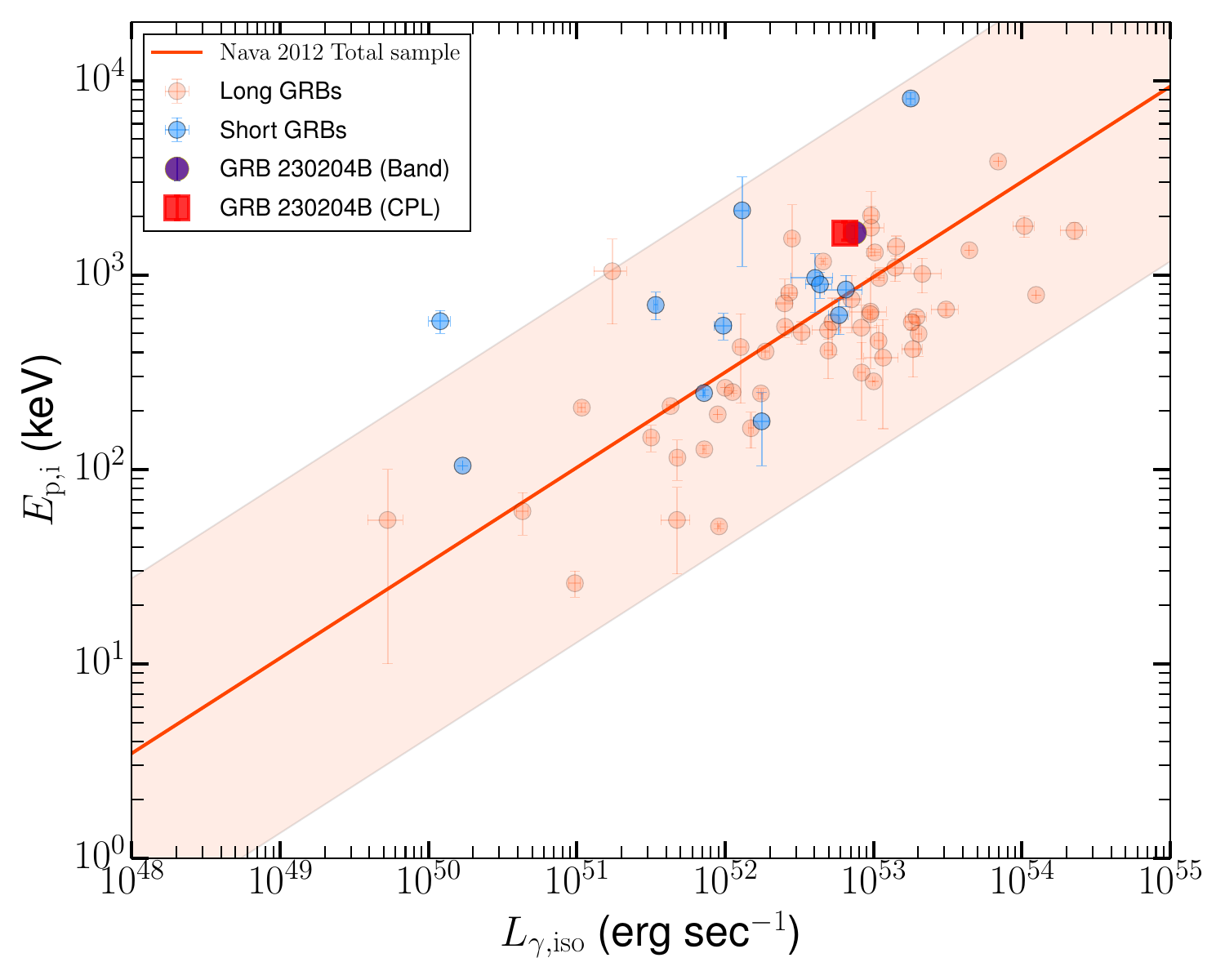} 
\includegraphics[scale=0.35]{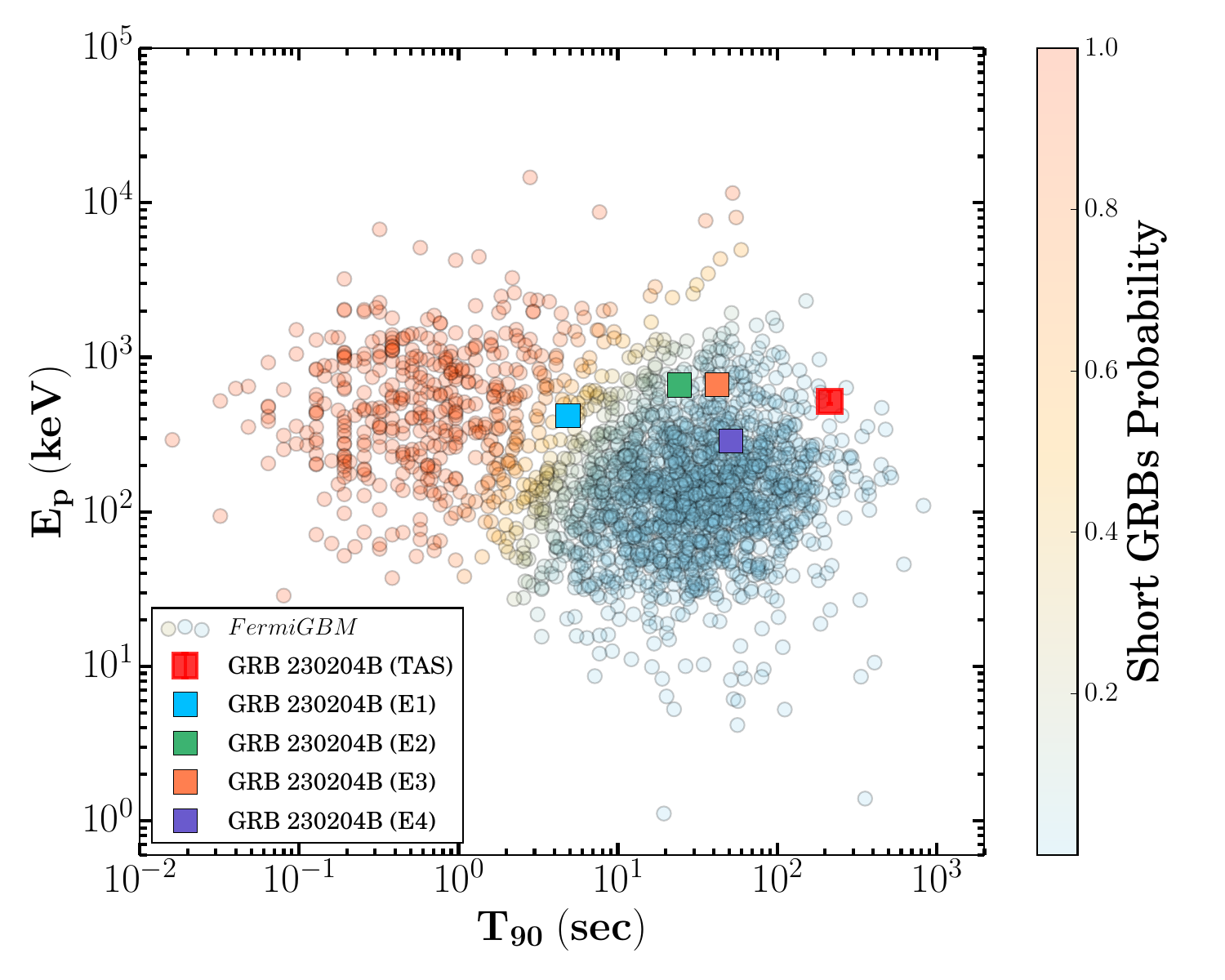} 
\caption{Top-left: Pulse-wise Amati correlation for \thisgrb. The position of individual pulses from GRB 230204B within the Amati correlation. Long and short GRBs, as extensively analyzed in \cite{2020MNRAS.492.1919M}, are represented by orange and blue circles, respectively, with corresponding solid lines indicating linear fits for these groups. The shaded regions surrounding the lines illustrate the 3$\sigma$ uncertainty bands. Top-right: The placement of GRB 230204B on the Yonetoku correlation. Comparative positions of well-studied long and short bursts from \cite{2012MNRAS.421.1256N} are marked by blue and orange circles, with parallel shaded bands signifying the 3$\sigma$ range. Both long and short GRBs follow the same plane for the Yonetoku correlation. Bottom: The pulse-wise \Ep-\tninty plane for \thisgrb and compare with a larger dataset of GRBs detected by the \fermi GBM.}
\label{fig:promptcorrelation}
\end{figure}

\begin{figure*}
\centering
\includegraphics[scale=0.5]{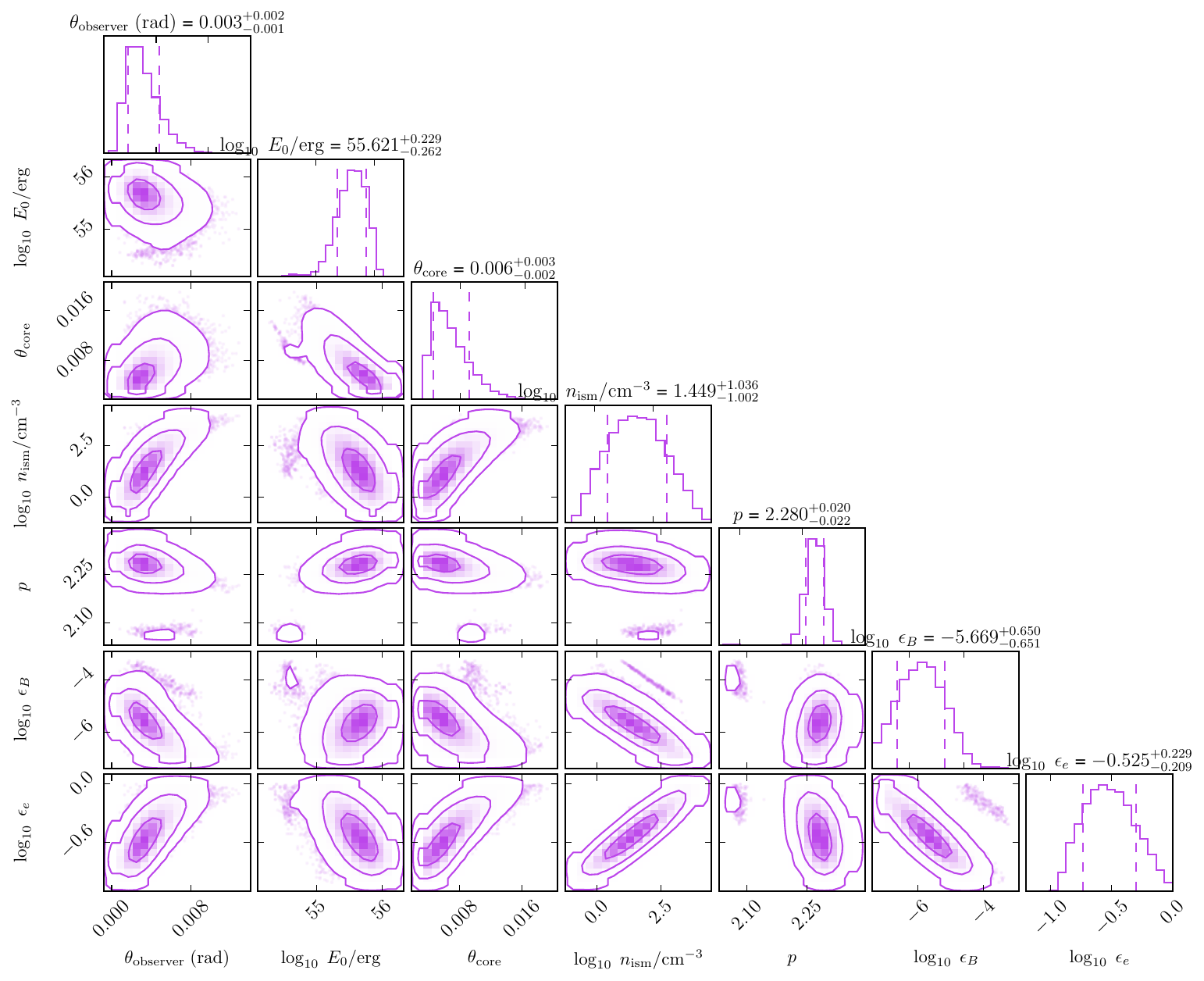} 
\caption{The corner plot presents the posterior distributions of the key afterglow model parameters, including jet opening angle, kinetic energy, and circumburst density. These distributions highlight the constraints derived from the broadband dataset and the robustness of the best-fit parameters.}
\label{fig:afterglowmodelcorner}
\end{figure*}

\begin{figure*}
\centering
\includegraphics[scale=0.4]{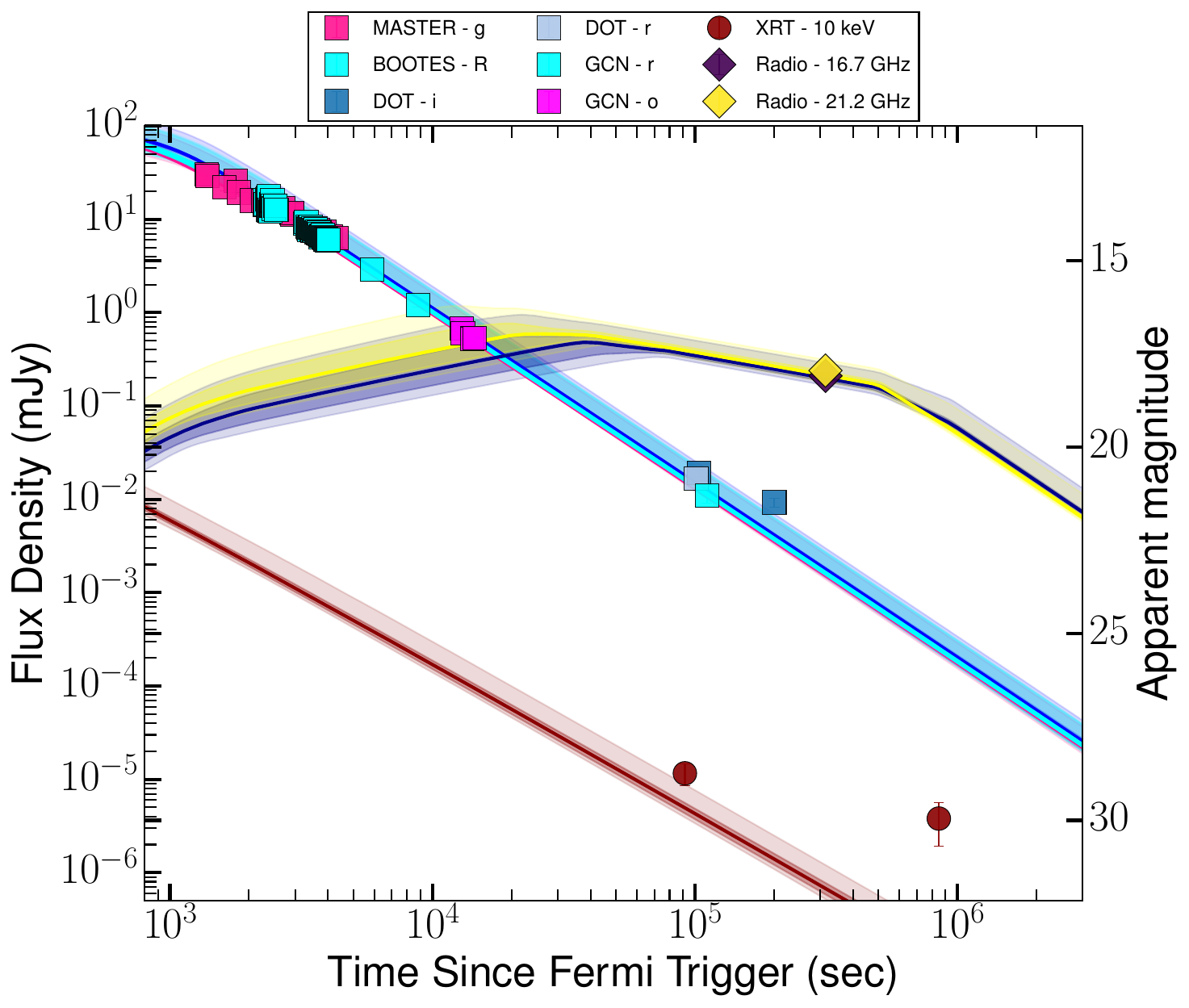} 
\caption{Same as Figure \ref{fig:afterglowmodel}, but with the light curve modeled using a top-hat forward shock jet scenario in a wind-like circumburst medium. The solid line represents the best-fit model, while the light and dark shaded regions correspond to the 1$\sigma$ and 2$\sigma$ confidence intervals, respectively.}
\label{fig:afterglowmodelwind}
\end{figure*}

\begin{table*}[!ht]
\centering
\caption{Physical parameters of the afterglow for GRB 230204B, derived through MCMC fitting using the \texttt{VegasAfterglow} framework for a wind-like environment. We have fixed the bulk Lorentz factor ~800 and $\theta_{\mathrm{v}}$ $\sim$ 0 rad.}
\label{afterglowpywind}
\begin{tabular}{lccc}\hline				
\textbf{Model}&\textbf{Top hat}&\textbf{Gaussian}&\textbf{Power-law}\\
\hline 
$\theta_{\mathrm{core}}~(\mathrm{rad})$ & 0.008$^{+0.001}_{-0.001}$ &  0.016$^{+0.002}_{-0.002}$ & 0.006$^{+0.001}_{-0.001}$ \\
$\log_{10}~E_{0}/{\mathrm{erg}}$ & 54.322$^{+0.099}_{-0.116}$& 54.087$^{+0.131}_{-0.153}$ & 54.620$^{+0.084}_{-0.085}$ \\
$p$ & 2.125$^{+0.021}_{-0.020}$ & 2.070$^{+0.045}_{-0.023}$  & 2.249$^{+0.026}_{-0.024}$ \\
$\log_{10}~A^{{*}}$ & -0.315$^{+0.150}_{-0.151}$& -0.069$^{+0.048}_{-0.079}$ & -0.221$^{+0.124}_{-0.146}$  \\
$\log_{10}~\epsilon_{B}$ & -2.168$^{+0.153}_{-0.139}$ & -3.066$^{+0.174}_{-0.145}$ &  -3.265$^{+0.228}_{-0.202}$\\
$\log_{10}~\epsilon_{e}$ & -1.376$^{+0.088}_{-0.078}$ & -0.684$^{+0.096}_{-0.141}$ &  -1.033$^{+0.071}_{-0.082}$\\
$\xi_{N}$& 1.0 & 1.0 & 1.0\\
\hline
\end{tabular}
\end{table*}

\begin{figure*}
\centering
\includegraphics[scale=0.4]{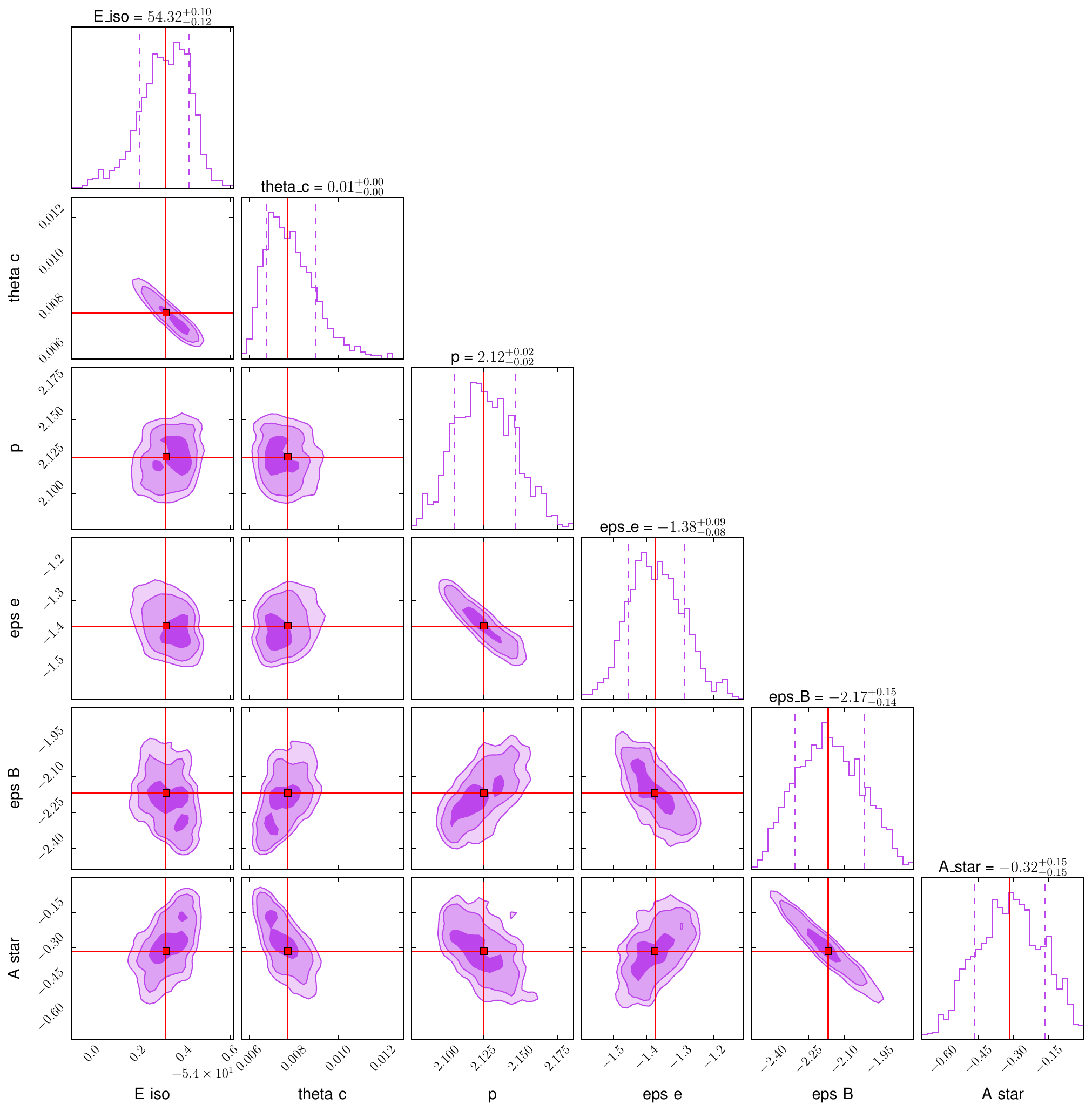} 
\caption{Similar as Fig. \ref{fig:afterglowmodelcorner} but for a wind-like circumburst medium.}
\label{fig:afterglowmodelcornerwind}
\end{figure*}

\begin{figure}
\centering
\includegraphics[scale=0.36]{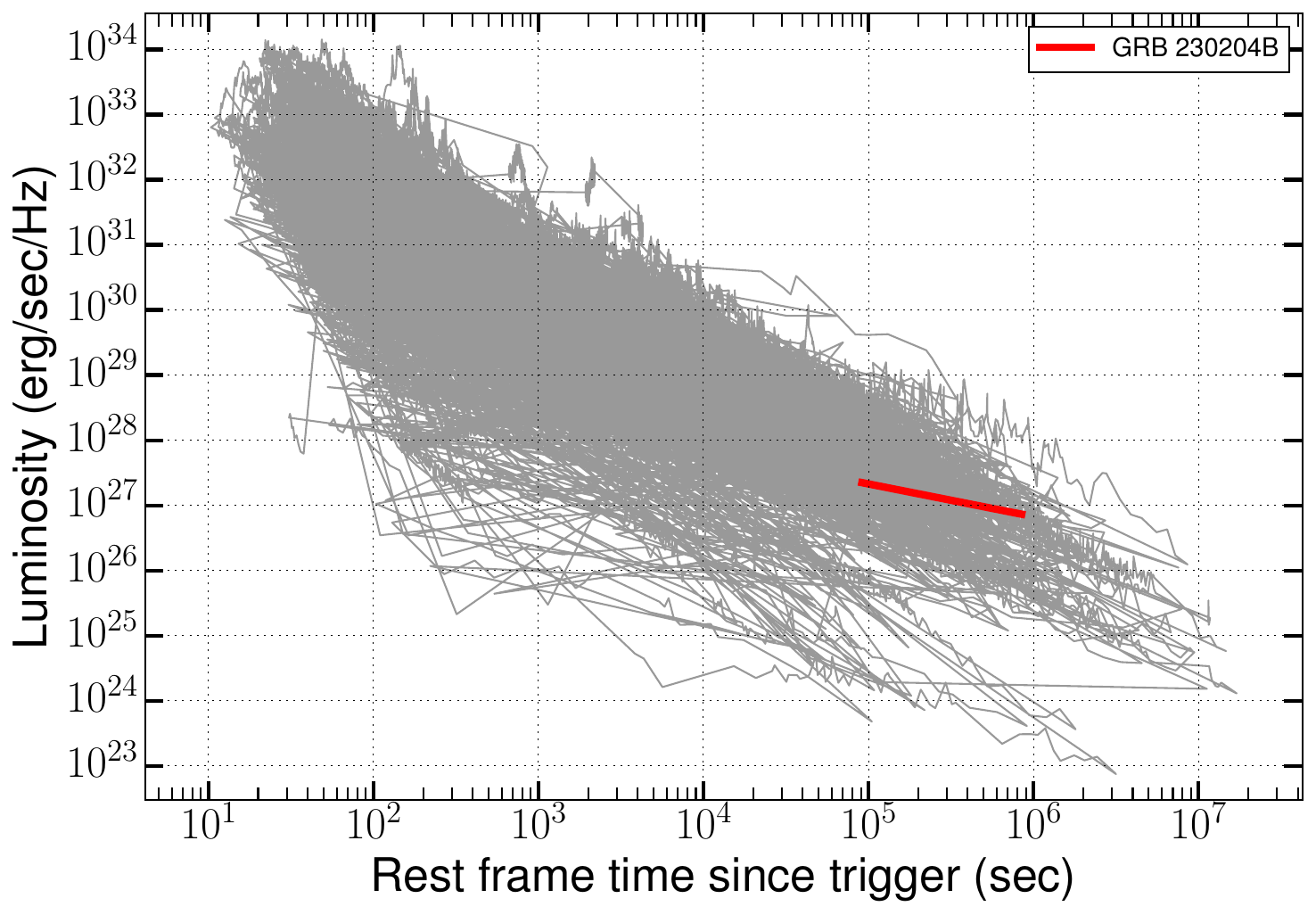} 
\includegraphics[scale=0.35]{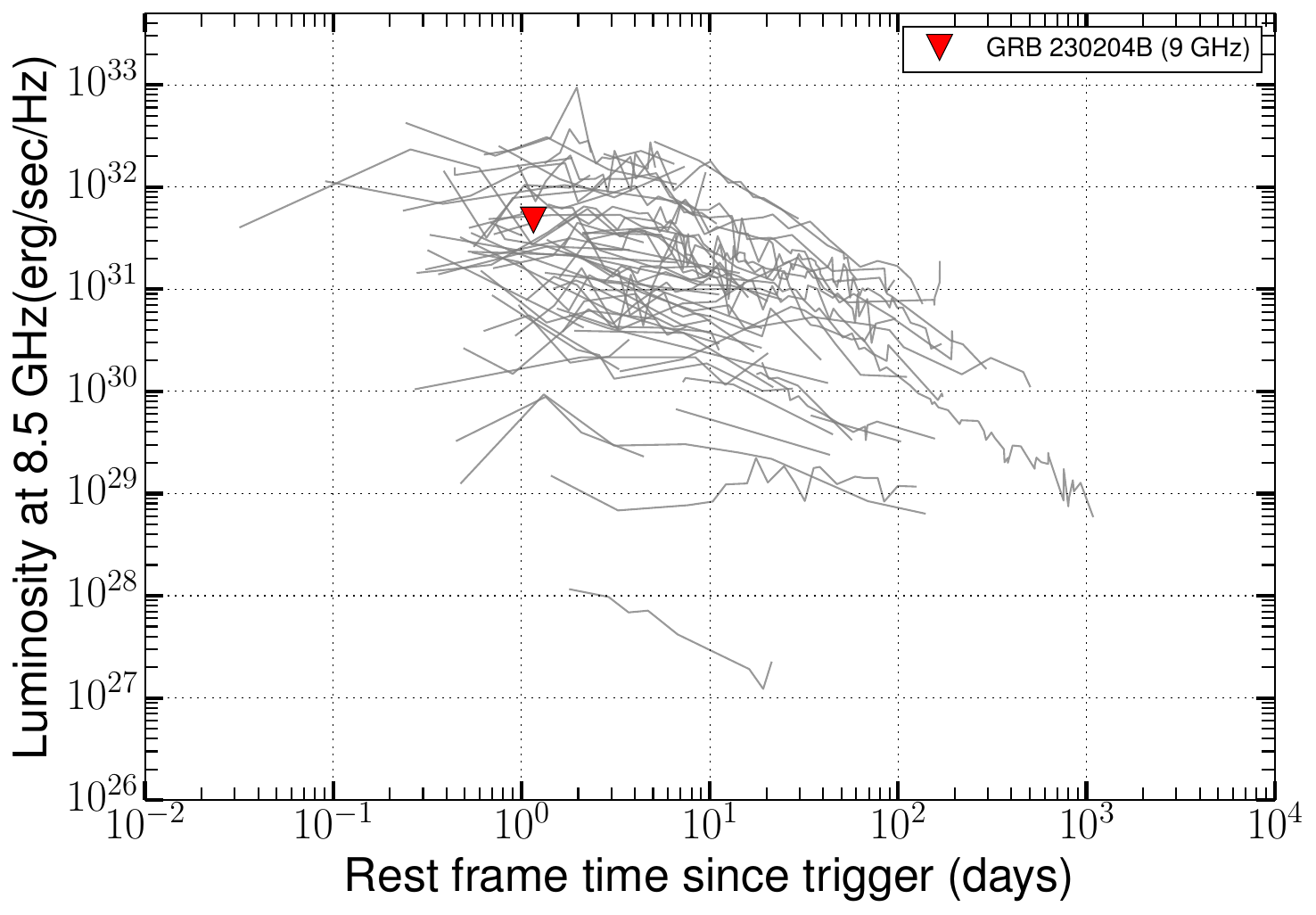} 
\caption{Left: X-ray luminosity light curves of GRB 230204B (in red) compared with other GRBs with known redshifts. The data were obtained from the \swift/XRT online repository \citep{2007A&A...469..379E, 2009MNRAS.397.1177E}. GRB 230204B's light curve is limited to two late-time data points, indicating an average luminosity consistent with other GRBs at similar epochs. The comparison underscores the alignment of GRB 230204B's late-phase luminosity with the general behavior observed in the broader GRB population. Right: Distribution of rest-frame 8.5 GHz radio luminosity light curves for a sample of GRBs with known redshifts (Shilling et al., in prep), where data points with SNR $<$ 2 are excluded for clarity. For \thisgrb, the 9 GHz upper limit observed with ATCA, taken several days post-burst, is overplotted for comparison, indicating a luminosity consistent with the general population. This placement highlights \thisgrb's relative position within the observed radio luminosity range of GRBs.}
\label{fig:radio}
\end{figure}

\begin{rotatetable}
\begin{tiny}
\begin{center}
\begin{longtable}{|c|c|ccccc|cccc|c|}
\caption{Time-resolved spectral fitting results for the all emission episode of GRB 230204B, analyzed using the \sw{Band} and \sw{CPL} models. The table includes key spectral parameters such as the low-energy spectral index ($\alpha_{\rm pt}$), high-energy spectral index ($\beta_{\rm pt}$) for the Band model, and the peak energy (\Ep) for both models. Flux values (in erg), computed within the energy range of 8 keV to 40 MeV, are provided for each time bin.} 
\label{TRS_Table}\\
\hline
T$_{\rm start}$ & T$_{\rm stop}$ & $\alpha_{\rm pt}$ & $\beta_{\rm pt}$ & \Ep &  Flux   & DIC$_{\rm Band}$ & $\Gamma_{\rm CPL}$ & $E_{\rm c}$ &  Flux  & DIC$_{\rm CPL}$ & \rm $\Delta$ DIC\\ \hline
 (sec) & (sec) &  & & (\keV) &  $\times 10^{-06}$ & &  &  (\keV) & $\times 10^{-06}$ &  & \\ \hline
-0.8 & 1.234 & $-0.31_{-0.21}^{0.23}$ & $-2.32_{-1.99}^{0.21}$ & $427.51_{-81.48}^{149.44}$ & 1.6 & 2080.12 & $-0.23_{-0.34}^{-0.01}$ & $252.01_{-0.63}^{188.35}$ & 1.31 & 2046.1 & 34.02 \\
1.234 & 4.973 & $-0.27_{-0.17}^{0.07}$ & $-2.21_{-1.42}^{0.09}$ & $333.22_{-17.26}^{87.52}$ & 2.54 & 2812.46 & $-0.38_{-0.11}^{0.05}$ & $252.35_{-20.10}^{47.30}$ & 1.76 & 2812.03 & 0.43 \\
41.412 & 45.259 & $-0.34_{-0.12}^{0.10}$ & $-4.70_{0.06}^{1.78}$ & $576.99_{-49.44}^{81.06}$ & 1.93 & 2844.64 & $-0.29_{-0.17}^{0.05}$ & $327.96_{-24.04}^{94.66}$ & 1.78 & 2828.86 & 15.78 \\
45.259 & 60.297 & $-0.61_{-0.05}^{0.02}$ & $-2.70_{-1.00}^{0.12}$ & $739.04_{-26.02}^{71.32}$ & 3.52 & 4361.92 & $-0.62_{-0.04}^{0.03}$ & $558.51_{-34.36}^{54.45}$ & 3.03 & 4369.86 & -7.94 \\
60.297 & 63.028 & $-0.83_{-0.10}^{0.11}$ & $-3.28_{-1.25}^{0.86}$ & $458.05_{-83.20}^{100.28}$ & 1.22 & 2501.94 & $-0.76_{-0.17}^{0.04}$ & $339.72_{-32.73}^{168.87}$ & 1.09 & 2502.2 & -0.26 \\
63.028 & 68.152 & $-0.82_{-0.30}^{0.12}$ & $-1.86_{-2.13}^{0.03}$ & $205.43_{-11.44}^{254.66}$ & 0.63 & 3138.76 & $-0.80_{-0.35}^{-0.01}$ & $214.89_{-0.64}^{412.66}$ & 0.42 & 3131.87 & 6.89 \\
109.451 & 122.927 & $-0.56_{-0.16}^{0.10}$ & $-4.08_{-0.52}^{1.32}$ & $534.81_{-52.64}^{129.77}$ & 0.71 & 4189.15 & $-0.46_{-0.25}^{0.02}$ & $321.61_{-10.84}^{174.97}$ & 0.64 & 4172.96 & 16.19 \\
122.927 & 140.987 & $-0.60_{-0.06}^{0.06}$ & $-3.49_{-1.07}^{0.58}$ & $854.61_{-84.42}^{102.21}$ & 1.9 & 4522.29 & $-0.59_{-0.07}^{0.05}$ & $607.82_{-62.14}^{117.27}$ & 1.79 & 4523.35 & -1.06 \\
140.987 & 155.493 & $-0.71_{-0.06}^{0.05}$ & $-4.50_{-0.14}^{1.42}$ & $637.62_{-63.90}^{72.21}$ & 1.56 & 4261.59 & $-0.70_{-0.07}^{0.04}$ & $480.27_{-45.15}^{95.07}$ & 1.47 & 4264.6 & -3.01 \\
155.493 & 156.992 & $-0.83_{-0.12}^{0.08}$ & $-2.45_{-2.09}^{-0.07}$ & $430.08_{-38.32}^{152.15}$ & 2.22 & 1898.35 & $-0.81_{-0.14}^{0.04}$ & $374.93_{-41.66}^{188.57}$ & 2.01 & 1900.76 & -2.41 \\
172.583 & 190.95 & $-0.94_{-0.06}^{0.04}$ & $-4.51_{-0.13}^{1.52}$ & $435.13_{-34.66}^{63.33}$ & 0.99 & 4436.83 & $-0.93_{-0.06}^{0.03}$ & $403.22_{-33.43}^{91.05}$ & 0.94 & 4443.64 & -6.81 \\
192.363 & 198.11 & $-0.83_{-0.08}^{0.21}$ & $-4.86_{0.36}^{2.50}$ & $273.23_{-68.82}^{39.77}$ & 0.63 & 3236.22 & $-0.73_{-0.20}^{0.04}$ & $195.33_{-14.73}^{100.71}$ & 0.57 & 3242.3 & -6.08 \\
198.11 & 200.125 & $-0.80_{-0.12}^{0.17}$ & $-4.21_{-0.40}^{1.40}$ & $228.30_{-39.58}^{37.12}$ & 0.82 & 2116.74 & $-0.67_{-0.24}^{0.03}$ & $157.75_{-9.14}^{87.55}$ & 0.77 & 2114.43 & 2.31 \\
200.125 & 230.0 & $-0.69_{-0.09}^{0.11}$ & $-4.75_{0.18}^{2.03}$ & $224.19_{-26.60}^{22.69}$ & 0.39 & 4862.37 & $-0.66_{-0.13}^{0.05}$ & $164.72_{-13.22}^{42.84}$ & 0.37 & 4862.75 & -0.38 \\ \hline
\end{longtable}
\end{center}
\end{tiny}
\end{rotatetable}

\clearpage

\begin{longtable}{|cccccc|}
\caption{Log of X-ray afterglow observations of GRB 230204B. These data were obtained from the \swift XRT repository.}
\label{tab:xrt}\\ \hline
Time-T0 (sec) & Energy (keV) & FD (uJy) & FD Err & Telescope & Ref. \\ \hline

\endfirsthead
\hline 
\multicolumn{6}{|c|}{{Continued on next page}} \\ \hline
\endfoot
\hline

Time-T0 (sec) & Energy (keV) & FD (uJy) & FD Err & Telescope & Ref. \\ \hline
\endhead
\hline
\endlastfoot
91528.12$^{+18774.81}_{-10700.53}$ & 10.0 & 0.0116 & 0.0030 & XRT & This work \\
848730.52$^{+17095.84}_{-7421.36}$ & 10.0 & 0.0038 & 0.0019 & XRT & This work \\
\end{longtable}

\begin{longtable}{|cccccc|}
\caption{Log of optical photometric observations of GRB 230204B.}
\label{tab:optical}\\ \hline
Time-T0 (sec) & Filter & MAG (AB) & MAG Err & Telescope & Ref. \\ \hline

\endfirsthead
\hline 
\multicolumn{6}{|c|}{{Continued on next page}} \\ \hline
\endfoot
\hline

Time-T0 (sec) & Filter & MAG (AB) & MAG Err & Telescope & Ref. \\ \hline
\endhead
\hline
\endlastfoot

1378.42698 & unfiltered & 12.7 & 0.12 & MASTER & This work \\
1777.40663 & unfiltered & 12.87 & 0.11 & MASTER & This work \\
2930.5762 & unfiltered & 13.77 & 0.17 & MASTER & This work \\
1389.55696 & unfiltered & 12.74 & 0.1 & MASTER & This work \\
1611.76601 & unfiltered & 13.03 & 0.1 & MASTER & This work \\
1830.95002 & unfiltered & 13.18 & 0.1 & MASTER & This work \\
2048.95398 & unfiltered & 13.4 & 0.11 & MASTER & This work \\
2267.91196 & unfiltered & 13.41 & 0.08 & MASTER & This work \\
2486.30301 & unfiltered & 13.54 & 0.07 & MASTER & This work \\
2694.05396 & unfiltered & 13.6 & 0.07 & MASTER & This work \\
2914.54193 & unfiltered & 13.74 & 0.08 & MASTER & This work \\
3654.57493 & unfiltered & 14.2 & 0.07 & MASTER & This work \\
3877.80896 & unfiltered & 14.23 & 0.04 & MASTER & This work \\
4097.35897 & unfiltered & 14.36 & 0.04 & MASTER & This work \\
4316.58904 & unfiltered & 14.41 & 0.04 & MASTER & This work \\ \hline
2296.5 & Clear & 13.51 & 0.06 & BOOTES & This work \\
2300.8 & Clear & 13.4 & 0.29 & BOOTES & This work \\
2319.4 & Clear & 13.5 & 0.13 & BOOTES & This work \\
2345.0 & Clear & 13.37 & 0.13 & BOOTES & This work \\
2363.8 & Clear & 13.33 & 0.17 & BOOTES & This work \\
2369.8 & Clear & 13.65 & 0.16 & BOOTES & This work \\
2375.9 & Clear & 13.29 & 0.16 & BOOTES & This work \\
2382.0 & Clear & 13.66 & 0.14 & BOOTES & This work \\
2399.4 & Clear & 13.65 & 0.14 & BOOTES & This work \\
2421.5 & Clear & 13.7 & 0.15 & BOOTES & This work \\
2427.6 & Clear & 13.52 & 0.13 & BOOTES & This work \\
2433.9 & Clear & 13.5 & 0.11 & BOOTES & This work \\
2439.9 & Clear & 13.55 & 0.28 & BOOTES & This work \\
2446.0 & Clear & 13.53 & 0.21 & BOOTES & This work \\
2459.1 & Clear & 13.4 & 0.1 & BOOTES & This work \\
2470.6 & Clear & 13.55 & 0.19 & BOOTES & This work \\
2491.9 & Clear & 13.62 & 0.2 & BOOTES & This work \\
2517.2 & Clear & 13.54 & 0.15 & BOOTES & This work \\
2523.4 & Clear & 13.55 & 0.18 & BOOTES & This work \\
2545.4 & Clear & 13.65 & 0.14 & BOOTES & This work \\
3269.6 & Clear & 14.02 & 0.06 & BOOTES & This work \\
3308.6 & Clear & 13.98 & 0.14 & BOOTES & This work \\
3335.6 & Clear & 14.1 & 0.08 & BOOTES & This work \\
3360.6 & Clear & 14.14 & 0.14 & BOOTES & This work \\
3396.6 & Clear & 14.18 & 0.08 & BOOTES & This work \\
3421.6 & Clear & 14.13 & 0.06 & BOOTES & This work \\
3446.6 & Clear & 14.1 & 0.09 & BOOTES & This work \\
3472.6 & Clear & 14.18 & 0.06 & BOOTES & This work \\
3497.6 & Clear & 14.2 & 0.08 & BOOTES & This work \\
3523.6 & Clear & 14.15 & 0.09 & BOOTES & This work \\
3548.6 & Clear & 14.19 & 0.04 & BOOTES & This work \\
3574.6 & Clear & 14.21 & 0.09 & BOOTES & This work \\
3599.6 & Clear & 14.16 & 0.08 & BOOTES & This work \\
3626.6 & Clear & 14.25 & 0.08 & BOOTES & This work \\
3651.6 & Clear & 14.23 & 0.08 & BOOTES & This work \\
3677.6 & Clear & 14.27 & 0.13 & BOOTES & This work \\
3703.6 & Clear & 14.25 & 0.08 & BOOTES & This work \\
3729.6 & Clear & 14.3 & 0.07 & BOOTES & This work \\
3755.6 & Clear & 14.38 & 0.09 & BOOTES & This work \\
3781.6 & Clear & 14.35 & 0.13 & BOOTES & This work \\
3806.6 & Clear & 14.35 & 0.08 & BOOTES & This work \\
3831.6 & Clear & 14.33 & 0.05 & BOOTES & This work \\
3858.6 & Clear & 14.38 & 0.07 & BOOTES & This work \\
3884.6 & Clear & 14.4 & 0.1 & BOOTES & This work \\
3909.6 & Clear & 14.44 & 0.06 & BOOTES & This work \\
3935.6 & Clear & 14.44 & 0.09 & BOOTES & This work \\
3960.6 & Clear & 14.46 & 0.05 & BOOTES & This work \\
3988.6 & Clear & 14.48 & 0.07 & BOOTES & This work \\
4028.6 & Clear & 14.45 & 0.08 & BOOTES & This work \\ 
27405.6 & Clear & $>$ 19.6 & -- & BOOTES & This work \\ 
36280.3 & $z$ & $>$ 17.7 & -- & BOOTES & This work \\ 
37487.6 & Clear & $>$ 19.7 & -- & BOOTES & This work \\ 
37558.0 & $i$ & $>$ 19.4 & -- & BOOTES & This work \\ 
72995.8 & $r$ & $>$ 19.3 & -- & BOOTES & This work \\ 
74544.8 & $i$ & $>$ 19.4 & -- & BOOTES & This work \\ \hline
101119.0 & $r$ &  21.09 & 0.09 & DOT & This work\\
278325.4 & $r$ & $>$ 21.27 & --&DOT & This work\\
103451.8 & $i$ & 20.88 & 0.10 & DOT & This work\\
200738.2 & $i$ & 21.67 & 0.12 & DOT & This work\\
105611.8 & $g$ & $>$ 21.4 & --& DOT & This work\\\hline
95569.7 & $V$ & $>$18.94 &  & UVOT & This work \\
95159.7 & $B$ & $>$19.50 &  & UVOT & This work \\
95075.4 & $U$ & $>$20.57 & & UVOT & This work \\
93868.0 & $UVW1$ & $>$19.53 &  & UVOT & This work \\
89647.6 & $UVM2$ & $>$21.7 &  & UVOT & This work \\
89126.7 & $UVW2$ & $>$19.78 & & UVOT & This work \\
841905.2 & $UVW2$ & $>$20.81 &  & UVOT & This work \\ \hline
5889.9 & $r$ & 15.25 & 0.04 & GIT & GCN \\
8801.5 & $r$ & 16.20 & 0.05 & GIT & GCN \\
12905.8 & $o$ & 16.82 & 0.04 & ATLAS & GCN \\
13064.8 & $o$ & 16.95 & 0.05 & ATLAS & GCN \\
14204.8 & $o$ & 17.10 & 0.05 & ATLAS & GCN \\
14477.8 & $o$ & 17.09 & 0.05 & ATLAS & GCN \\
110948.8 & $r$ & 21.30 & 0.18 & VLT & GCN \\ \hline
\end{longtable}

\begin{table}
\centering
\caption{Log of radio follow-up observations of GRB 230204B.}
\begin{tabular}{|cccccc|}\hline
Time-T0 (sec) & Freq (GHz) & FD (mJy) & FD Err & Telescope & Ref. \\ \hline
314133 & 5.5 & $<$0.45 &  & ATCA & This work \\
314133 & 9.0 & $<$0.20 &  & ATCA & This work \\
314133 & 16.7 & 0.21 & 0.02 & ATCA & This work \\
314133 & 21.2 & 0.24 & 0.05 & ATCA & This work \\
314133 & 33.0 & $<$0.12 & & ATCA & This work \\
314133 & 35.0 & $<$0.13 &  & ATCA & This work \\
12658049 & 0.842 & $<$1.31 &  &ASKAP  & This work \\
28948293 & 0.943 & $<$1.45 &  &ASKAP & This work \\\hline
\end{tabular}
\label{tab:radio}
\end{table}

\begin{table}
\centering
\caption{MAXI constrained (3 $\sigma$) on the X-ray afterglow emission of GRB 230204B.}
\begin{tabular}{|ccc|}\hline
T-T$_{0}$ (sec) & Energy range exposure & Upper limit (mJy) \\ \hline
 22526 & 2-20 keV & 0.052 \\
 28102 & 2-20 keV & 0.042 \\
 33679 & 2-20 keV & 0.047 \\
 39257 & 2-20 keV & 0.047 \\
 44834 & 2-20 keV & 0.058 \\
 50409 & 2-20 keV & 0.062 \\
 55986 & 2-20 keV & 0.058 \\
 61563 & 2-20 keV & 0.057 \\
 67140 & 2-20 keV & 0.053 \\
 72717 & 2-20 keV & 0.055 \\
 78292 & 2-20 keV & 0.055
 \\ \hline
\end{tabular}
\label{tab:MAXI}
\end{table}

\bibliography{GRB230204B}{}
\bibliographystyle{aasjournal}

\end{document}